\documentclass[aps,preprint]{revtex4}%
\usepackage{amsfonts}
\usepackage{amsmath}
\usepackage{amssymb}
\usepackage{graphicx}%
\newtheorem{theorem}{Theorem}
\newtheorem{acknowledgement}[theorem]{Acknowledgement}

\begin{document}
\title[A further study of $p$-adic open string amplitudes]{ $p$-Adic open string amplitudes with Chan-Paton factors coupled to a constant
$B$-field }
\author{H. Garc\'{\i}a-Compe\'{a}n}
\affiliation{Centro de Investigaci\'{o}n y de Estudios Avanzados del Instituto
Polit\'{e}cnico Nacional, Departamento de F\'{\i}sica,P.O. Box 14-740, CP.
07000, M\'{e}xico D.F., M\'{e}xico.}
\email{compean@fis.cinvestav.mx, elopez@fis.cinvestav.mx}
\thanks{The work of E. Y. L\'{o}pez was supported by a CONACyT fellowship.}
\author{Edgar Y. L\'{o}pez}
\affiliation{Centro de Investigaci\'{o}n y de Estudios Avanzados del Instituto
Polit\'{e}cnico Nacional, Departamento de F\'{\i}sica,P.O. Box 14-740, CP.
07000, M\'{e}xico D.F., M\'{e}xico.}
\author{W. A. Z\'{u}\~{n}iga-Galindo}
\affiliation{Centro de Investigaci\'{o}n y de Estudios Avanzados del Instituto
Polit\'{e}cnico Nacional, Departamento de Matem\'{a}ticas, Unidad
Quer\'{e}taro, Libramiento Norponiente \#2000, Fracc. Real de Juriquilla.
Santiago de Quer\'{e}taro, Qro. 76230, M\'{e}xico.}
\email{wazuniga@math.cinvestav.edu.mx}
\thanks{This author was partially supported CONACyT grant 250845.}

\begin{abstract}
We establish rigorously the regularization of the $p$-adic open string
amplitudes, with Chan-Paton rules and a constant $B$-field, introduced by
Ghoshal and Kawano. In this study we use techniques of multivariate local zeta
functions depending on multiplicative characters and a phase factor which
involves an antisymmetric bilinear form. These local zeta functions are new
mathematical objects. We attach to each amplitude a multivariate local zeta
function depending on the kinematic parameters, the $B$-field and the
Chan-Paton factors. We show that these integrals admit meromorphic
continuations in the kinematic parameters. This result allows us to regularize
the Ghoshal-Kawano amplitudes. The regularized amplitudes do not have
ultraviolet divergencies. Due to the need for a certain symmetry, the theory
works only for prime numbers which are congruent to $3$ modulo $4$. We also
discuss the limit $p\rightarrow1$ in the noncommutative effective field theory
and in the Ghoshal-Kawano amplitudes. We show that in the case of four points,
the limit $p\rightarrow1$ of the regularized Ghoshal-Kawano amplitudes
coincides with the Feynman amplitudes attached to the limit $p\rightarrow1$ of
the noncommutative Gerasimov-Shatashvili Lagrangian.

\end{abstract}
\date[10/09/2019]{}
\maketitle
\tableofcontents

\section{Introduction}

The deep connections between $p$-adic analysis and physics are a natural
consequence of the emergence of ultrametricity in physics, which means the
occurrence of ultrametric spaces in physical models, see e.g. \cite{Dragovivh
et al,Vol,Volovich1,Ramal et
al,V-V-Z,Khrennikov,KKZuniga,Varadarajan,Zuniga-LNM-2016} and the references
therein. The existence of a Planck length implies that the spacetime
considered as a topological space is completely disconnected. The points
(which are the connected components) play the role of spacetime quanta. This
is precisely the Volovich conjecture on the non-Archimedean nature of the
spacetime below the Planck scale, \cite{Vol,Volovich1}, \cite[Chapter
6]{Varadarajan}. Ultrametric spaces have also appeared in models of complex
systems. A central paradigm in the physics of \ certain complex systems (for
instance proteins), asserts that the dynamics of such systems can be modeled
as a random walk over the leaves of a rooted tree. This tree is a finite
ultrametric space constructed out of the energy landscape. Mean-field
approximations of these models drive naturally to models involving $p$-adic
numbers, see e.g. \cite{KKZuniga,Zun-JPA,Zun-Nonlinearity,Zuni-ANOA}, and the
references therein.

In the last forty years, the above mentioned ideas have motivated many
developments in quantum field theory and string theory, see e.g.
\cite{Dragovivh et al,Hlousek:1988vu,Brekke:1993gf,V-V-Z}, and more recently,
\cite{Gubser:2016guj,Heydeman:2016ldy,Gubser:2016htz,Dutta:2017bja,Gubser:2017qed,Mendoza-Martinez:2018ktr,Jepsen:2018dqp,Stoica:2018zmi,Gubser:2018cha,Heydeman:2018qty,Hung:2018mcn,Parikh:2019ygo,Huang:2019nog,Hung:2019zsk}%
.

In string theory, the scattering amplitudes are obtained integrating over the
moduli space of Riemann surfaces. Even for tree-level amplitudes (on the
sphere for closed strings and on the disk for open strings) $N$-point
amplitudes are difficult to compute beyond four points. Moreover, the
convergence of these integrals is not evident by itself
\cite{Berera:1992tm,Witten:2013pra}. Recently, in \cite{ZunigaBVeys} was
established in a rigorous mathematical way that Koba-Nielsen amplitudes are
bona fide integrals, which admit meromorphic continuations when considered as
complex functions of the kinematic parameters.

String theory with a $p$-adic world-sheet was proposed and studied for the
first time in \cite{Freund:1987kt}. Later this theory was formally known as
$p$-adic string theory. The Adelic scattering amplitudes which are related to
the Archimedean ones were studied in \cite{Freund:1987ck}. The tree-level
string amplitudes were explicitly computed in the case of $p$-adic string
world-sheet in \cite{Brekke:1988dg} and \cite{Frampton:1987sp}. These
amplitudes can be formally obtained from a suitable action using general
principles \cite{Spokoiny:1988zk}. A general treatment, starting by describing
a discrete field theory on a Bruhat-Tits tree and obtaining the tree-level
string amplitudes (\cite{Brekke:1988dg}), was established in
\cite{Zabrodin:1988ep}. Similarly as in the standard string theory, in
$p$-adic string theories it is difficult to determine the convergence region
in the momentum space, however this was done precisely for the $N$-point tree
amplitudes in \cite{Bocardo-Gaspar:2016zwx}. In this article we show (in a
rigorous mathematical way) that the $p$-adic open string $N-$point tree
amplitudes are bona fide integrals that admit meromorphic continuations as
rational functions, by relating them with multivariate local zeta functions
(also called multivariate Igusa local zeta functions \cite{Igusa, Meuser,
Denef}).

In $p$-adic string theory the limit $p\rightarrow1$ is very intriguing since
it seems to be related to the real versions of these theories
\cite{Gerasimov:2000zp,Spokoiny:1988zk}. This limit is special since the
effective theory shows that it is related to physical string theories such as
the boundary string field theory \cite{Witten:1992qy}. Another interpretation
of the limit $p\rightarrow1$ was given in terms of the renormalization group
scaling transformation in the Bruhat-Tits tree for some suitable $p$
\cite{Ghoshal:2006te}. In the worldsheet theory we cannot forget the nature of
$p$ as a prime number, thus the analysis of the limit is more subtle. The
correct way of taking the limit $p\rightarrow1$ involves the introduction of
finite extensions of the $p$-adic field $\mathbb{Q}_{p}$. In
\cite{Bocardo-Gaspar:2017atv} the limit $p\rightarrow1$ was discussed at the
tree-level string amplitudes. We provided a rigorous definition of this limit
using the theory of topological zeta functions due to Denef and Loeser
\cite{denefandloeser,trossmann}.

In ordinary string theory the effective action for bosonic open strings in
gauge field backgrounds was discussed many years ago in
\cite{Abouelsaood:1986gd}. The analysis incorporating a Neveu-Schwarz $B$
field in the target space leads to a noncommutative effective gauge theory on
the world-volume of D-branes \cite{Seiberg:1999vs}. The study of the $p$-adic
open string tree amplitudes including Chan-Paton factors was started in
\cite{Brekke:1988dg}. However the incorporation of a $B$-field in the $p$-adic
context and the computation of the tree level string amplitudes was discussed
in \cite{Ghoshal:2004ay,Grange:2004xj}. In these works it was reported that
the tree-level string amplitudes are affected by a noncommutative factor. In
\cite{Ghoshal:2004ay} Ghoshal and Kawano introduced new amplitudes involving
multiplicative characters and a noncommutative factor. These amplitudes
coincide with the ones obtained directly from the noncommutative effective
action \cite{Ghoshal:2004dd}.

In the present article we study the $p$-adic string amplitudes, with
Chan-Paton rules and a constant $B$-field, introduced in \cite{Ghoshal:2004ay}%
, by using techniques of multivariate local zeta functions. \ The $N$-point,
$p$-adic, open string amplitudes, with Chan-Paton rules in a constant
$B$-field, have the form
\begin{equation}
{\int\limits_{\mathbb{Q}_{p}^{N}}}\text{ }{\prod\limits_{1\leq i<j\leq N}%
}\left\vert x_{i}-x_{j}\right\vert _{p}^{\boldsymbol{k}_{i}\boldsymbol{k}_{j}%
}H_{\tau}\left(  x_{i}-x_{j}\right)  \text{ }\exp\left\{  -\frac{\sqrt{-1}}%
{2}\left(  {\sum\limits_{1\leq i<j\leq N}}(\boldsymbol{k}_{i}\theta
\boldsymbol{k}_{j})\mathrm{sgn}_{\tau}(x_{i}-x_{j})\right)  \right\}
{\prod\limits_{i=1}^{N}}dx_{i}\text{,} \label{Amplitude}%
\end{equation}
where $N\geq4$, $\boldsymbol{k}=\left(  \boldsymbol{k}_{1},\ldots
,\boldsymbol{k}_{N}\right)  $, $\boldsymbol{k}_{i}=\left(  k_{0,i}%
,\ldots,k_{l,i}\right)  $, $i=1,\ldots,N$, is the momentum vector of the
$i$-th tachyon vertex operator (with Minkowski product $\boldsymbol{k}%
_{i}\boldsymbol{k}_{j}=-k_{0,i}k_{0,j}+k_{1,i}k_{1,j}+\cdots+k_{l,i}k_{l,j}$)
obeying
\begin{equation}
\sum_{i=1}^{N}\boldsymbol{k}_{i}=\boldsymbol{0}\text{, \ \ \ \ }%
\boldsymbol{k}_{i}\boldsymbol{k}_{i}=2\text{ \ for }i=1,\ldots,N,
\label{momentumconservation}%
\end{equation}
$H_{\tau}\left(  x_{i}\right)  =\frac{1}{2}(1+\mathrm{sgn}_{\tau}(x))$,
$\mathrm{sgn}_{\tau}(x)$ is a $p$-adic version of the sign function, $\theta$
is a fixed antisymmetric bilinear form, and ${\prod\nolimits_{i=1}^{N}}dx_{i}$
is the normalized Haar measure of $\mathbb{Q}_{p}^{N}$. A symmetry requirement
for this theory is that the sign function be odd, meaning $sgn_{\tau
}(-x)=-sgn_{\tau}(x)$.

Unfortunately, this theory is not invariant under projective M\"{o}bius
transformations and consequently the normalization $x_{1}=0$, $x_{N-1}=1$,
$x_{N}=\infty$ can not be carried out. This is a consequence of the fact that
$\mathbb{Q}_{p}$ is not an ordered field. Anyway, in \cite{Ghoshal:2004ay} the
authors assumed a such normalization, which is equivalent to assuming that the
vertex operators are inserted in the boundary of the Bruhat-Tits tree at
`non-generic points'. Taking the normalization $x_{1}=0$, $x_{N-1}=1$,
$x_{N}=\infty$, the amplitude takes the form
\begin{align}
A^{(N)}\left(  \boldsymbol{k},\theta,\tau\right)   &  ={\int
\limits_{\mathbb{Q}_{p}^{N-3}}}{\prod\limits_{i=2}^{N-2}}\left\vert
x_{i}\right\vert _{p}^{\boldsymbol{k}_{1}\boldsymbol{k}_{i}}\left\vert
1-x_{i}\right\vert _{p}^{\boldsymbol{k}_{N-1}\boldsymbol{k}_{i}}H_{\tau
}\left(  x_{i}\right)  H_{\tau}\left(  1-x_{i}\right) \nonumber\\
&  \times\text{ }{\prod\limits_{2\leq i<j\leq N-2}}\left\vert x_{i}%
-x_{j}\right\vert _{p}^{\boldsymbol{k}_{i}\boldsymbol{k}_{j}}H_{\tau}\left(
x_{i}-x_{j}\right) \nonumber\\
&  \times\text{ }\exp\left\{  -\frac{\sqrt{-1}}{2}\left(  {\sum\limits_{1\leq
i<j\leq N-1}}(\boldsymbol{k}_{i}\theta\boldsymbol{k}_{j})\mathrm{sgn}_{\tau
}(x_{i}-x_{j})\right)  \right\}  {\prod\limits_{i=2}^{N-2}}dx_{i}\text{.}
\label{amplituduno}%
\end{align}
We have called such integrals Ghoshal-Kawano amplitudes. The main goal of this
article is the study of the amplitude $A^{(N)}\left(  \boldsymbol{k}%
,\theta,\tau\right)  $ using twisted multivariate Igusa's local zeta
functions. We attach to $A^{(N)}\left(  \boldsymbol{k},\theta,\tau\right)  $
the following Igusa type integral:
\begin{gather}
Z^{(N)}\left(  \boldsymbol{s},\widetilde{\boldsymbol{s}},\tau\right)
={\int\limits_{\mathbb{Q}_{p}^{N-3}}}{\prod\limits_{i=2}^{N-2}}\left\vert
x_{i}\right\vert _{p}^{s_{1i}}\left\vert 1-x_{i}\right\vert _{p}^{s_{\left(
N-1\right)  i}}H_{\tau}\left(  x_{i}\right)  H_{\tau}\left(  1-x_{i}\right)
\nonumber\\
\times\text{ }{\prod\limits_{2\leq i<j\leq N-2}}\left\vert x_{i}%
-x_{j}\right\vert _{p}^{s_{ij}}H_{\tau}\left(  x_{i}-x_{j}\right)  \text{
}\exp\left\{  -\frac{\sqrt{-1}}{2}\left(  {\sum\limits_{1\leq i<j\leq N-1}%
}\widetilde{s}_{ij}\mathrm{sgn}_{\tau}(x_{i}-x_{j})\right)  \right\}
{\prod\limits_{i=2}^{N-2}}dx_{i}\text{,} \label{F_1}%
\end{gather}
where the $s_{ij}$ are complex symmetric variables and the $\widetilde{s}%
_{ij}$ are real antisymmetric variables. We have called integrals
$Z^{(N)}\left(  \boldsymbol{s},\widetilde{\boldsymbol{s}},\tau\right)  $
Ghoshal-Kawano local zeta functions. As a consequence of the presence of the
Chan-Paton factors, the normalization $x_{1}=0$, $x_{N-1}=1$, $x_{N}=\infty$
and the requirement that $sgn_{\tau}(-x)=-sgn_{\tau}(x)$, the integration in
(\ref{F_1}) is actually on $\mathbb{Z}_{p}^{N-3}$ (the $N-3$-dimensional unit
ball). This fact implies that turning off the background $B$-field, the
\ amplitude $A^{(N)}\left(  \boldsymbol{k},\theta,\tau\right)  $ does not
reduce to the $p$-adic open string amplitude at the tree level. This fact was
already noticed in \cite{Ghoshal:2004ay}, in the case $N=4$. We show that
integrals $Z^{(N)}(\boldsymbol{s},\widetilde{\boldsymbol{s}},\tau)$ can be
expressed as finite sums of a new type of twisted multivariate Igusa's local
zeta functions, and by using the techniques of \cite{Igusa,Loeser}, we
establish that $Z^{\left(  N\right)  }(\boldsymbol{s},\widetilde
{\boldsymbol{s}},\tau)$ admits a meromorphic continuation as a rational
function in the variables $p^{-s_{1j}}$, $p^{-s_{\left(  N-1\right)  j}}$,
\ $p^{-s_{ij}}$. Furthermore, $Z^{\left(  N\right)  }(\boldsymbol{s}%
,\widetilde{\boldsymbol{s}},\tau)$ is holomorphic in
\begin{equation}
{\bigcap\limits_{\mathcal{H}}}\left\{  s_{ij}\in\mathbb{C}^{\mathbf{d}};\text{
}\sum_{i,j\in M}N_{ij,k}\operatorname{Re}\left(  s_{ij}\right)  +\gamma
_{k}>0,\text{ for }k\in T\right\}  , \label{Holo-Region}%
\end{equation}
where $N_{ij,k}\in\mathbb{N}$, $\gamma_{k},\mathbf{d}\in
\mathbb{N\smallsetminus}\left\{  0\right\}  $, and $M$, $T$ are finite sets,
and the real parts of its poles belong to the finite union of hyperplanes of
type
\[
\mathcal{H}=\left\{  s_{ij}\in\mathbb{C}^{\mathbf{d}};\text{ }\sum_{ij\in
M}N_{ij,k}\operatorname{Re}\left(  s_{ij}\right)  +\gamma_{k}=0,\text{ for
}k\in T\right\}  .
\]
We regularize the amplitude $A^{\left(  N\right)  }\left(  \boldsymbol{k}%
,\theta,\tau\right)  $ by redefining it as
\[
A^{\left(  N\right)  }\left(  \boldsymbol{k},\theta,\tau\right)  =Z^{\left(
N\right)  }(\boldsymbol{s},\widetilde{\boldsymbol{s}},\tau)\mid
_{\substack{s_{ij}=\boldsymbol{k}_{i}\boldsymbol{k}_{j}\\\widetilde{s}%
_{ij}=\boldsymbol{k}_{i}\theta\boldsymbol{k}_{j}}},
\]
in this way $A^{\left(  N\right)  }\left(  \boldsymbol{k},\theta,\tau\right)
$ is a well-defined meromorphic function of the kinematic parameters
$\boldsymbol{k}_{i}\boldsymbol{k}_{j}$, which agrees with integral
(\ref{amplituduno}), if it exists. As a consequence of the description of the
poles of $Z^{\left(  N\right)  }(\boldsymbol{s},\widetilde{\boldsymbol{s}%
},\tau)$, $A^{\left(  N\right)  }\left(  \boldsymbol{k},\theta,\tau\right)  $
is defined for arbitrary large momenta, since in (\ref{Holo-Region}) the
values $\boldsymbol{k}_{i}\boldsymbol{k}_{j}$ can take arbitrarily large
values. This fact is not valid for the $p$-adic Koba-Nielsen amplitudes, see
\cite{Bocardo-Gaspar:2016zwx}, and \cite{ZunigaBVeys}, since Ghoshal-Kawano
amplitudes are supposed to be generalizations of the $p$-adic open amplitudes
at the tree level, we conclude that the normalization $x_{1}=0$, $x_{N-1}=1$,
$x_{N}=\infty$ is not possible in the presence of a background $B$-field. In a
forthcoming article we expect to study amplitudes (\ref{Amplitude}). It is
worth to mention here that the Ghoshal-Kawano local zeta functions are new
Igusa-type integrals coming from $p$-adic string theory.

The construction of a physical theory over a $p-$adic spacetime (worldsheet in
our case) raises the question about the physical meaning of the prime $p$. The
spacetime is a quadratic space $(\mathbb{Q}_{p}^{N},\mathfrak{q})$, where
$\mathfrak{q}$ is a quadratic form, and consequently, the spacetime depends on
the pair $\left(  p,\mathfrak{q}\right)  $. In this article, we require
$p\equiv3$ $\operatorname{mod}4$ and $\tau\in\{p,\varepsilon p\}$ in order to
have the symmetry $\mathrm{sgn}_{\tau}(-x)=-\mathrm{sgn}_{\tau}(x)$.

This article is organized as follows. In section \ref{Section_II}, we study
the limit $p\rightarrow1$ in the noncommutative version of the effective
action discussed in \cite{Ghoshal:2004dd}. We describe the noncommutative
version of the Gerasimov-Shatashvili action and find explicitly its four-point
amplitudes. In section \ref{Section_III}, we review the basic aspects of the
twisted, multivariate Igusa's local zeta functions. The local zeta functions
required here are a variation of the ones considered in \cite{Loeser}.
Sections \ref{Section_IV}-\ref{Section_V} are dedicated to establish the
meromorphic continuation of Ghoshal-Kawano local zeta functions. Sections
\ref{Section_VI} and \ref{Section_VII} are devoted to give the explicit
calculation for the $4$-point and $5$-point amplitudes. The $4$-point
amplitude was already obtained by Ghoshal and Kawano in \cite{Ghoshal:2004ay}
under certain hypotheses and the $5$-point amplitude is new. In section
\ref{Section_VIII}, we compute the $p\rightarrow1$ limit of the $p$-adic
$4$-point and $5$-point amplitudes. We verified that the $p\rightarrow1$ limit
of $4$-point amplitude coincides with the Feynman amplitude computed from the
noncommutative Gerasimov-Shatashvili action in section \ref{Section_II}. The
final remarks are collected in section \ref{Section_IX}. Finally in the
Appendix, we review the basic aspects of the $p$-adic analysis, and introduce
some notations and conventions used along this article.

\section{\label{Section_II}The limit $p\rightarrow1$ in the effective action
with a $B$-field}

\subsection{\label{Section_II_A}The limit $p\rightarrow1$ in the
noncommutative effective action}

In \cite{Ghoshal:2004dd}, it was considered a noncommutative action as the
effective action of the theory of $p$-adic open strings with a $B$-field. The
corresponding action in the $D$-dimensional spacetime is given by
\begin{equation}
S(\phi)={\frac{1}{g^{2}}}{\frac{p^{2}}{p-1}}\int d^{D}x\bigg(-{\frac{1}{2}%
}\phi\star p^{-{\frac{1}{2}}\Delta}\phi+{\frac{1}{p+1}}(\star\phi
)^{p+1}\bigg), \label{goodaction}%
\end{equation}
where $g$ and $\Delta$ are the coupling constant and the Laplacian,
respectively, and $(\star\phi)^{p}$ is defined by $\phi\star\phi\star
\cdots\star\phi$ $p$-times. Here $\star$ is the Moyal star product, which is
defined for any suitable pair of smooth functions $f$ and $g$ as
\[
{(f\star g)(x)=\exp\bigg({\frac{i}{2}}\theta^{\mu\nu}{\frac{\partial}{\partial
y^{\mu}}}{\frac{\partial}{\partial z^{\nu}}}\bigg)f(x+y)g(x+z)\bigg|_{y=z=0}.}%
\]
The corresponding equation of motion is given by
\begin{equation}
p^{-{\frac{1}{2}}\Delta}\phi=(\star\phi)^{p}. \label{eomgeneral}%
\end{equation}
The solutions of this equation are defined in the target space $\mathbb{R}%
^{D}$, where $p$ plays the role of a real parameter. In particular the limit
$p$ approaches to one makes sense.

Now following \cite{Gerasimov:2000zp}, by considering the Taylor expansion of
$\exp(-\frac{1}{2}\Delta\log p)$ and $\exp(p\log(\star\phi))$ at $p=1$, and
keeping only the linear term, we get
\begin{equation}
\Delta\phi=-2\phi\star\log(\star\phi), \label{eomlineal}%
\end{equation}
where $\log(\star\phi)=\phi-\frac{1}{2}\phi\star\phi+\frac{1}{3}\phi\star
\phi\star\phi-\cdots$. Thus the heuristic $p\rightarrow1$ limit leads to a
noncommutative version of the Gerasimov and Shatashvili Lagrangian:
\begin{equation}
S(\phi)=\int d^{D}x\bigg((\partial\phi)^{2}-V(\star\phi)\bigg),
\label{linearaction}%
\end{equation}
where
\[
V(\phi)=(\star\phi)^{2}\star\log\bigg[{\frac{(\star\phi)^{2}}{e}}\bigg].
\]

In noncommutative field theory, it is well known that the nontrivial
noncommutative effect comes from the potential energy of the Lagrangian. The
propagators associated with the kinetic energy of the Lagrangian are the same
as the ones of the commutative theory. Thus the free Lagrangian with an
external source $J(x)$ is
\[
S_{0}(\phi)=\int d^{D}x\big[(\partial\phi)^{2}+\phi^{2}(x)+J(x)\phi(x)\big].
\]
The propagators are given by $x_{ij}=\frac{1}{{\boldsymbol{k}_{i}%
\cdot\boldsymbol{k}_{j}}+1},$ where $k_{i}$, with $i=1,\dots,N$, are the
external momenta of the particles. The Feynman rule for the interaction vertex
can be obtained in the noncommutative theory by considering the cubic,
quartic, etc. interaction terms and computing the correlation functions, see
for instance, \cite{Minwalla:1999px,Micu:2000xj}.

\subsection{\label{Section_II_B}Amplitudes from the noncommutative
Gerasimov-Shatashvili Lagrangian}

In this subsection we show how to extract the four-point amplitudes from the
noncommutative Gerasimov-Shatashvili Lagrangian (\ref{linearaction}). In order
to do that, we first require to study the interacting theory. The generating
functional of the correlation function for the free theory is given by
\[
\mathcal{Z}_{0}[J]=\mathcal{N}[\det(\Delta-1)]^{-1/2}\exp\bigg\{-{\frac
{i}{2\hbar}}\int d^{D}x\int d^{D}x^{\prime}J(x)G_{F}(x-x^{\prime})J(x^{\prime
})\bigg\},
\]
where $G_{F}(x-x^{\prime})$ is the Green function of time-ordered product of
two fields of the theory, $\mathcal{N}$ is a normalization constant,
$[\det(\Delta-1)]^{-1/2}$ is a suitable regularization of the divergent
determinant bosonic operator. The noncommutative action is given by
\begin{equation}
S(\phi)=\int d^{D}x\big[(\partial\phi)^{2}+\phi^{2}-U(\star\phi)\big],
\label{actionU}%
\end{equation}
where $U(\star\phi)=2(\star\phi)^{2}\star\log(\star\phi)$. We expand
$U(\star\phi)$ in Taylor series as follows:
\begin{equation}
U(\star\phi)=A\phi\star\phi+B\phi\star\phi\star\phi+C\phi\star\phi\star
\phi\star\phi+\cdots, \label{potentialU}%
\end{equation}
where $A=-\frac{25}{6}$, $B=8$ and $C=-6$.

The generating $Z[J]$ functional incorporating the interaction is given by
\[
\mathcal{Z}[J]=\exp\bigg\{{\frac{25i}{6\hbar}}\int d^{D}x\bigg(-i\hbar
{\frac{\delta}{\delta J(x)}}\bigg)\star\bigg(-i\hbar{\frac{\delta}{\delta
J(x)}}\bigg)
\]%
\[
{-\frac{8i}{\hbar}}\int d^{D}x\bigg(-i\hbar{\frac{\delta}{\delta J(x)}%
}\bigg)\star\bigg(-i\hbar{\frac{\delta}{\delta J(x)}}\bigg)\star
\bigg(-i\hbar{\frac{\delta}{\delta J(x)}}\bigg)
\]%
\begin{equation}
+{\frac{6i}{\hbar}}\int d^{D}x\bigg(-i\hbar{\frac{\delta}{\delta J(x)}%
}\bigg)\star\bigg(-i\hbar{\frac{\delta}{\delta J(x)}}\bigg)\star
\bigg(-i\hbar{\frac{\delta}{\delta J(x)}}\bigg)\star\bigg(-i\hbar{\frac
{\delta}{\delta J(x)}}\bigg)+\cdots\bigg\}\mathcal{Z}_{0}[J].
\end{equation}
We are interested in checking whether connected tree-level scattering
amplitudes of this theory match exactly with the corresponding $p$-adic
amplitudes in the limit when $p$ tends to one. The computation of the field
theory performed here will be compared to the computation of the $p$-adic
string amplitudes at section \ref{Section_VIII}.

\subsection{\label{Section_II_C}Four-point amplitudes}

In this subsection we consider the quartic term from the potential
(\ref{potentialU}). The expansion of the exponential function of this term in
the interacting generating functional is expressed as
\[
\mathcal{Z}[J]=\cdots+6i\hbar^{3}\int d^{D}x\bigg\{\bigg({\frac{\delta}{\delta
J(x)}}\bigg)\star\bigg({\frac{\delta}{\delta J(x)}}\bigg)\star\bigg({\frac
{\delta}{\delta J(x)}}\bigg)\star\bigg({\frac{\delta}{\delta J(x)}%
}\bigg)\bigg\}\mathcal{Z}_{0}[J]+\cdots
\]%
\[
=\cdots+6i\hbar^{3}\lim_{x=y_{1}=y_{2}=y_{3}=y_{4}}\lim_{w_{1}=w_{2}%
=w_{3}=w_{4}=0}\int d^{D}y_{1}d^{D}y_{2}d^{D}y_{3}d^{D}y_{4}%
\]%
\[
\times\exp\bigg\{{\frac{i}{2}}\theta^{\mu_{1}\nu_{1}}{\frac{\partial}{\partial
w_{1}^{\mu_{1}}}}{\frac{\partial}{\partial w_{2}^{\nu_{1}}}}\bigg\}\exp
\bigg\{{\frac{i}{2}}\theta^{\mu_{2}\nu_{2}}{\frac{\partial}{\partial
w_{3}^{\mu_{2}}}}{\frac{\partial}{\partial w_{4}^{\nu_{2}}}}\bigg\}
\]%
\begin{equation}
\times\bigg\{\bigg({\frac{\delta}{\delta J(y_{1}+w_{1})}}\bigg)\bigg({\frac
{\delta}{\delta J(y_{2}+w_{2})}}\bigg)\bigg({\frac{\delta}{\delta
J(y_{3}+w_{3})}}\bigg)\bigg({\frac{\delta}{\delta J(y_{4}+w_{4})}%
}\bigg)\bigg\}\mathcal{Z}_{0}[J]+\cdots.
\end{equation}

A straightforward computation of the $4$-point vertex gives
\[
{\frac{\delta^{4}\mathcal{Z}[J]}{\delta J(x_{1})\delta J(x_{2})\delta
J(x_{3})\delta J(x_{4})}}\bigg|_{J=0}%
\]%
\[
=\cdots+{768i\hbar^{3}}\int d^{D}x\bigg\{\bigg\{\cos\bigg({\frac{\partial
_{1}\theta\partial_{2}}{2}}\bigg)\cos\bigg({\frac{\partial_{3}\theta
\partial_{4}}{2}}\bigg)+\cos\bigg({\frac{\partial_{1}\theta\partial_{3}}{2}%
}\bigg)\cos\bigg({\frac{\partial_{2}\theta\partial_{4}}{2}}\bigg)
\]%
\[
+\cos\bigg({\frac{\partial_{1}\theta\partial_{4}}{2}}\bigg)\cos\bigg({\frac
{\partial_{2}\theta\partial_{3}}{2}}\bigg)\bigg\}\bigg[-{\frac{i}{2\hbar}%
}G_{F}(x-x_{1})\bigg]\bigg[-{\frac{i}{2\hbar}}G_{F}(x-x_{2})\bigg]
\]%
\begin{equation}
\times\bigg[-{\frac{i}{2\hbar}}G_{F}(x-x_{3})\bigg]\bigg[-{\frac{i}{2\hbar}%
}G_{F}(x-x_{4})\bigg]+\cdots, \label{vertex}%
\end{equation}
where $G_{F}(x-y)$ is the propagator and $\partial_{1,2,3,4}$ are the partial
derivative with respect to the coordinates $x_{1}$, $x_{2}$, $x_{3}$ and
$x_{4}$, respectively.

The interaction term $8(\star\phi)^{3}$ in the Lagrangian gives also a
non-vanishing contribution to the 4-point tree amplitudes of the second order
in perturbation theory. They are described by Feynman diagrams with two
vertices located at points $y$ and $z$ connected by a propagator $G_{F}(y-z)$
and with two external legs attached to each vertex. In this case the amplitude
is computed from the relevant part of the generating functional:
\begin{equation}
\mathcal{Z}[J]=\cdots+{64\hbar^{4}}\int d^{D}y\int d^{D}z\ \bigg(\star
{\frac{\delta}{\delta J(y)}}\bigg)^{3}\bigg(\star{\frac{\delta}{\delta J(z)}%
}\bigg)^{3}\mathcal{Z}_{0}[J]+\cdots.\label{cubiccubic}%
\end{equation}
This expression can be written explicitly in terms of the Moyal product as
\[
\mathcal{Z}[J]=\cdots+{64\hbar^{4}}\lim_{y=y_{1}=y_{2}=y_{3}}\lim
_{z=z_{1}=z_{2}=z_{3}}\lim_{w_{1}=w_{2}=w_{3}=w_{4}=0}\int d^{D}y_{1}%
d^{D}y_{2}d^{D}y_{3}d^{D}z_{1}d^{D}z_{2}d^{D}z_{3}%
\]%
\[
\times\exp\bigg\{{\frac{i}{2}}\theta^{\mu_{1}\nu_{1}}{\frac{\partial}{\partial
w_{1}^{\mu_{1}}}}{\frac{\partial}{\partial w_{2}^{\nu_{1}}}}\bigg\}\exp
\bigg\{{\frac{i}{2}}\theta^{\mu_{2}\nu_{2}}{\frac{\partial}{\partial
w_{3}^{\mu_{2}}}}{\frac{\partial}{\partial w_{4}^{\nu_{2}}}}\bigg\}
\]%
\[
\times\bigg\{\bigg({\frac{\delta}{\delta J(y_{1}+w_{1})}}\bigg)\bigg({\frac
{\delta}{\delta J(y_{2}+w_{2})}}\bigg)\bigg({\frac{\delta}{\delta J(y_{3})}%
}\bigg)\bigg\}
\]%
\[
\times\bigg\{\bigg({\frac{\delta}{\delta J(z_{1}+w_{3})}}\bigg)\bigg({\frac
{\delta}{\delta J(z_{2}+w_{4})}}\bigg)\bigg({\frac{\delta}{\delta J(z_{3})}%
}\bigg)\bigg\}\mathcal{Z}_{0}[J]+\cdots.
\]
The connected 4-point amplitudes at the second order coming from the cubic
interaction $B\phi^{3}$ yields to
\[
{\frac{\delta^{4}\mathcal{Z}[J]}{\delta J(x_{1})\delta J(x_{2})\delta
J(x_{3})\delta J(x_{4})}}\bigg|_{J=0}=\cdots+8192\hbar^{4}\int d^{D}y\int
d^{D}z\ \bigg[-{\frac{i}{2\hbar}}G_{F}(y-z)\bigg]
\]%
\[
\times\bigg\{\cos\bigg({\frac{\partial_{1}\theta\partial_{2}}{2}}%
\bigg)\cos\bigg({\frac{\partial_{3}\theta\partial_{4}}{2}}\bigg)
\]%
\[
\times\bigg\{\bigg[-{\frac{i}{2\hbar}}G_{F}(y-x_{1})\bigg]\bigg[-{\frac
{i}{2\hbar}}G_{F}(y-x_{2})\bigg]\bigg[-{\frac{i}{2\hbar}}G_{F}(z-x_{3}%
)\bigg]\bigg[-{\frac{i}{2\hbar}}G_{F}(z-x_{4})\bigg]
\]%
\[
+\bigg[-{\frac{i}{2\hbar}}G_{F}(z-x_{1})\bigg]\bigg[-{\frac{i}{2\hbar}}%
G_{F}(z-x_{2})\bigg]\bigg[-{\frac{i}{2\hbar}}G_{F}(y-x_{3})\bigg]\bigg[-{\frac
{i}{2\hbar}}G_{F}(y-x_{4})\bigg]\bigg\}
\]%
\[
+\cos\bigg({\frac{\partial_{1}\theta\partial_{3}}{2}}\bigg)\cos\bigg({\frac
{\partial_{2}\theta\partial_{4}}{2}}\bigg)
\]%
\[
\times\bigg\{\bigg[-{\frac{i}{2\hbar}}G_{F}(y-x_{1})\bigg]\bigg[-{\frac
{i}{2\hbar}}G_{F}(z-x_{2})\bigg]\bigg[-{\frac{i}{2\hbar}}G_{F}(y-x_{3}%
)\bigg]\bigg[-{\frac{i}{2\hbar}}G_{F}(z-x_{4})\bigg]
\]%
\[
+\bigg[-{\frac{i}{2\hbar}}G_{F}(z-x_{1})\bigg]\bigg[-{\frac{i}{2\hbar}}%
G_{F}(y-x_{2})\bigg]\bigg[-{\frac{i}{2\hbar}}G_{F}(z-x_{3})\bigg[-{\frac
{i}{2\hbar}}G_{F}(y-x_{4})\bigg]\bigg\}
\]%
\[
+\cos\bigg({\frac{\partial_{1}\theta\partial_{4}}{2}}\bigg)\cos\bigg({\frac
{\partial_{2}\theta\partial_{3}}{2}}\bigg)
\]%
\[
\times\bigg\{\bigg[-{\frac{i}{2\hbar}}G_{F}(z-x_{1})\bigg]\bigg[-{\frac
{i}{2\hbar}}G_{F}(y-x_{2})\bigg]\bigg[-{\frac{i}{2\hbar}}G_{F}(y-x_{3}%
)\bigg]\bigg[-{\frac{i}{2\hbar}}G_{F}(z-x_{4})\bigg]
\]%
\begin{equation}
+\bigg[-{\frac{i}{2\hbar}}G_{F}(y-x_{1})\bigg]\bigg[-{\frac{i}{2\hbar}}%
G_{F}(z-x_{2})\bigg]\bigg[-{\frac{i}{2\hbar}}G_{F}(z-x_{3})\bigg[-{\frac
{i}{2\hbar}}G_{F}(y-x_{4})\bigg]\bigg\}\bigg\}+\cdots.\label{canalesstu}%
\end{equation}
This total amplitude corresponds exactly to the sum of the partial amplitudes
associated to the channels $s$, $t$ and $u$. The sum of (\ref{canalesstu}) and
(\ref{vertex}) constitutes the $4$-point amplitude (at the tree-level). This
amplitude agrees with the limit $p\rightarrow1$ of the sum over the
permutations of the momenta $\boldsymbol{k}_{i}$ of the  $4$-point $p$-adic
amplitudes computed in section \ref{Section_VI}. The details of this
calculation are given in section \ref{Section_VIII}. Moreover, five-point
non-commutative amplitudes (and higher-order amplitudes) in the limit
$p\rightarrow1$ can be computed following a similar procedure, but it will not
be performed here.

\section{\label{Section_III}Multivariate Local Zeta Functions}

For the notation and the definition of basic objects such as multiplicative
characters, sign functions, Haar measure, etc., the reader may consult the
Appendix. In this section we review some basic aspects of the twisted
multivariate local zeta functions. The meromorphic continuation of the local
zeta functions play a central role in sections \ref{Section_IV} and
\ref{Section_V}.

Let $f_{1}(x),\ldots,f_{m}(x)\in\mathbb{Q}_{p}\left[  x_{1},\ldots
,x_{n}\right]  $ be non-constant polynomials, we denote by $\mathbb{D}%
:=\cup_{i=1}^{m}f_{i}^{-1}(0)$ the divisor attached to them. Let $\chi
_{1},\ldots,\chi_{m}$ be multiplicative characters. We set $\boldsymbol{f}%
:=\left(  f_{1},\ldots,f_{m}\right)  \text{, \ }\boldsymbol{\chi}:=\left(
\chi_{1},\ldots,\chi_{m}\right)  \text{,\ \ and \ }\boldsymbol{s}:=\left(
s_{1},\ldots,s_{m}\right)  \in\mathbb{C}^{m}\text{.}$ The multivariate local
zeta function attached to $(\boldsymbol{f},\boldsymbol{\chi},\Theta)$, with
$\Theta\mathcal{\ }$a test function (i.e. a locally constant function with
compact support), is defined as
\begin{equation}
Z_{\Theta}\left(  \boldsymbol{s},\boldsymbol{\chi},\boldsymbol{f}\right)
=\int\limits_{\mathbb{Q}_{p}^{n}\smallsetminus\mathbb{D}}\Theta\left(
x\right)  \prod\limits_{i=1}^{m}\chi_{i}\left(  ac\left(  f_{i}(x)\right)
\right)  \left\vert f_{i}(x)\right\vert _{p}^{s_{i}}{\prod\limits_{i=1}^{n}%
}dx_{i}\text{, } \label{Zeta
Function general}%
\end{equation}
with $\operatorname{Re}(s_{i})>0$ for all $i$. Integrals of type
(\ref{Zeta Function general}) are holomorphic functions in $\boldsymbol{s}$,
which admit meromorphic continuations to the whole $\mathbb{C}^{m}$,
\cite[Th\'{e}or\`{e}me 1.1.4.]{Loeser}, see also \cite{Igusa}. More precisely,
the integrals $Z_{\Theta}\left(  \boldsymbol{s},\boldsymbol{\chi
},\boldsymbol{f}\right)  $ admit meromorphic continuations as rational
functions in the variables $p^{-s_{1}},\ldots,p^{-s_{m}}$. Let us emphasize
that the notation $\chi_{i}\left(  ac(x)\right)  $, $x\neq0$, means that the
character $\chi_{i}$ depends only on the angular component of $x$, see Appendix.

We need a variation of the Loeser result \cite[Th\'{e}or\`{e}me 1.1.4.]%
{Loeser}, more precisely, when each $\chi_{i}\circ ac$ is the trivial
character $\chi_{\text{triv}}\left(  x\right)  $ or $\mathrm{sgn}_{\tau
}\left(  x\right)  $. We denote by $\chi_{i}$ one of these characters. This
last function is a multiplicative character on $\mathbb{Q}_{p}^{\times}$, but
it depends on the angular component of $x$ and on the order of $x$. By using
Hironaka's resolution of singularities theorem, see e.g. \cite{Igusa},
$Z_{\Theta}\left(  \boldsymbol{s},\boldsymbol{\chi},\boldsymbol{f}\right)  $
can be written as linear combination of integrals of type
\[
{\int\limits_{c+p^{e}\mathbb{Z}_{p}^{n}}}\text{ }{\prod\limits_{j=1}^{r}%
}\left\{  \left\vert y_{j}\right\vert _{p}^{\sum_{i=1}^{m}N_{i,j}s_{i}%
+v_{j}-1}\chi_{i}^{N_{i,j}}\left(  y_{j}\right)  \right\}  dy_{j},
\]
where $c=\left(  c_{1},\ldots,c_{n}\right)  \in\mathbb{Q}_{p}^{n}$, $1\leq
r\leq n$, $N_{i,j}$ are nonnegative integers for $i\in\left\{  1,\ldots
,m\right\}  $, $j\in\mathcal{T}$, and $v_{j}$ a positive integer, for
$j\in\mathcal{T}$ (a finite set), see proof of \cite[Theorem 8.2.1]{Igusa} and
\cite[Th\'{e}or\`{e}me 1.1.4.]{Loeser}.

Then, we have to study the meromorphic continuation of an integral of type%
\[
I(s):={\int\limits_{c_{j}+p^{e}\mathbb{Z}_{p}}}\text{ }\left\vert
y_{j}\right\vert _{p}^{\sum_{i=1}^{m}N_{i,j}s_{i}+v_{j}-1}\mathrm{sgn}_{\tau
}^{N_{i,j}}\left(  y_{j}\right)  dy_{j},
\]
since the one corresponding to the trivial character is already known, see
e.g. \cite[Lemma 8.2.1]{Igusa}. Several cases occur. If $c_{j}\notin
p^{e}\mathbb{Z}_{p}$, by using the fact that $\left\vert \cdot\right\vert
_{p}$ and $\mathrm{sgn}_{\tau}\left(  \cdot\right)  $ are locally constant
functions we get that%
\[
I(s)=p^{-e}\text{ }\left\vert c_{j}\right\vert _{p}^{\sum_{i=1}^{m}%
N_{i,j}s_{i}+v_{j}-1}\mathrm{sgn}_{\tau}^{N_{i,j}}\left(  c_{j}\right)  .
\]
In the case $c_{j}\in p^{e}\mathbb{Z}_{p}$, we have%
\begin{align*}
I(s)  &  ={\sum\limits_{l=e}^{\infty}}\text{ }{\int\limits_{p^{l}%
\mathbb{Z}_{p}^{\times}}}\text{ }\left\vert y_{j}\right\vert _{p}^{\sum
_{i=1}^{m}N_{i,j}s_{i}+v_{j}-1}\mathrm{sgn}_{\tau}^{N_{i,j}}\left(
y_{j}\right)  dy_{j}\\
&  =\left\{  {\sum\limits_{l=e}^{\infty}}\text{ }p^{-l\left(  \sum_{i=1}%
^{m}N_{i,j}s_{i}+v_{j}\right)  }\mathrm{sgn}_{\tau}^{N_{i,j}}\left(
p^{l}\right)  \right\}  {\int\limits_{\mathbb{Z}_{p}^{\times}}}\text{
}\mathrm{sgn}_{\tau}^{N_{i,j}}\left(  u\right)  du\\
&  =:J(s){\int\limits_{\mathbb{Z}_{p}^{\times}}}\text{ }\mathrm{sgn}_{\tau
}^{N_{i,j}}\left(  u\right)  du,
\end{align*}
where $y_{j}=p^{l}u$.

Now if $\tau=\varepsilon$, $\mathrm{sgn}_{\tau}\left(  u\right)  =\left(
-1\right)  ^{ord(u)}\equiv1$ for any $u\in\mathbb{Z}_{p}^{\times}$, then
$\int_{\mathbb{Z}_{p}^{\times}}$ $\mathrm{sgn}_{\varepsilon}^{N_{i,j}}\left(
u\right)  du$ $=1-p^{-1}$. In the case $\tau\neq\varepsilon$,
\[
{\int\limits_{\mathbb{Z}_{p}^{\times}}}\text{ }\mathrm{sgn}_{\tau}^{N_{i,j}%
}\left(  u\right)  du=\left\{
\begin{array}
[c]{lll}%
1-p^{-1} & \text{if } & N_{i,j}\text{ is even}\\
&  & \\
0 & \text{if} & N_{i,j}\text{ is odd.}%
\end{array}
\right.
\]
By using that
\[
\mathrm{sgn}_{\tau}^{N_{i,j}}\left(  p^{l}\right)  =\mathrm{sgn}_{\tau
}^{lN_{i,j}}\left(  p\right)  =\left\{
\begin{array}
[c]{lll}%
1 & \text{if } & l\text{ is even}\\
&  & \\
\mathrm{sgn}_{\tau}^{N_{i,j}}\left(  p\right)  & \text{if} & l\text{ is odd,}%
\end{array}
\right.
\]
we have
\begin{align*}
J(s)  &  ={\sum\limits_{l=e}^{\infty}}\text{ }p^{-l\left(  \sum_{i=1}%
^{m}N_{i,j}s_{i}+v_{j}\right)  }\mathrm{sgn}_{\tau}^{lN_{i,j}}\left(  p\right)
\\
&  ={\sum\limits_{k=0}^{\infty}}\text{ }p^{-\left(  k+e\right)  \left(
\sum_{i=1}^{m}N_{i,j}s_{i}+v_{j}\right)  }\mathrm{sgn}_{\tau}^{N_{i,j}}\left(
p^{k+e}\right) \\
&  =p^{-e\left(  \sum_{i=1}^{m}N_{i,j}s_{i}+v_{j}\right)  }\mathrm{sgn}_{\tau
}^{N_{i,j}}\left(  p^{e}\right)  {\sum\limits_{k=0}^{\infty}}\text{
}p^{-k\left(  \sum_{i=1}^{m}N_{i,j}s_{i}+v_{j}\right)  }\mathrm{sgn}_{\tau
}^{kN_{i,j}}\left(  p\right)  .
\end{align*}
If $N_{f_{i},j}$ is even
\begin{align*}
J(s)  &  =p^{-e\left(  \sum_{i=1}^{m}N_{i,j}s_{i}+v_{j}\right)  }%
{\sum\limits_{k=0}^{\infty}}\text{ }p^{-k\left(  \sum_{i=1}^{m}N_{i,j}%
s_{i}+v_{j}\right)  }\\
&  =\frac{p^{-e\left(  \sum_{i=1}^{m}N_{i,j}s_{i}+v_{j}\right)  }}%
{1-p^{-\sum_{i=1}^{m}N_{i,j}s_{i}-v_{j}}}.
\end{align*}
If $N_{i,j}$ is \ odd, we have $I(s)=0$. In conclusion, since $Z_{\Theta
}\left(  \boldsymbol{s},\boldsymbol{\chi},\boldsymbol{f}\right)  $ is a finite
linear combination \ of products of integrals of type $I(s)$, then $Z_{\Theta
}\left(  \boldsymbol{s},\boldsymbol{\chi},\boldsymbol{f}\right)  $ \ admits a
meromorphic continuation as a rational function in the variables $p^{-s_{1}%
},\ldots,p^{-s_{m}}$. More precisely,%
\begin{equation}
Z_{\Theta}\left(  \boldsymbol{s},\boldsymbol{\chi},\boldsymbol{f}\right)
=\frac{L_{\Theta,\boldsymbol{\chi}}\left(  \boldsymbol{s}\right)  }%
{\prod\limits_{j\in\mathcal{T}}\left(  1-p^{-\sum_{i=1}^{m}N_{i,j}s_{j}-v_{j}%
}\right)  }, \label{Multizeta-special-case}%
\end{equation}
where $L_{\Theta,\boldsymbol{\chi}}\left(  \boldsymbol{s}\right)  $ is a
polynomial in the variables $p^{-s_{1}},\ldots,p^{-s_{m}}$, and the real parts
of its poles belong to the finite union of hyperplanes
\[
{\sum\limits_{i=1}^{m}}N_{i,j}\operatorname{Re}\left(  s_{i}\right)
+v_{j}=0,\quad\text{ for }j\in\mathcal{T}.
\]
This result is a variation of \cite[Th\'{e}or\`{e}me 1.1.4.]{Loeser}.

\section{\label{Section_IV}The Ghoshal-Kawano local zeta function}

From now on, we use $\theta$ to denote a fixed antisymmetric bilinear form. In
\cite{Ghoshal:2004ay} Ghoshal and Kawano proposed the following amplitude (for
the $N$-point tree-level, $p$-adic open string amplitude, with Chan-Paton
rules in a constant $B$-field):
\begin{align*}
A^{(N)}\left(  \boldsymbol{k},\theta,\tau;x_{1},x_{N-1}\right)   &
:={\int\limits_{\mathbb{Q}_{p}^{N-3}\smallsetminus\mathbb{D}}}\text{{ }}%
{\prod\limits_{i=2}^{N-2}}\left\vert x_{i}\right\vert _{p}^{\boldsymbol{k}%
_{1}\boldsymbol{k}_{i}}\left\vert 1-x_{i}\right\vert _{p}^{\boldsymbol{k}%
_{N-1}\boldsymbol{k}_{i}}H_{\tau}\left(  x_{i}\right)  H_{\tau}\left(
1-x_{i}\right) \\
&  \times\text{ }{\prod\limits_{2\leq i<j\leq N-2}}\left\vert x_{i}%
-x_{j}\right\vert _{p}^{\boldsymbol{k}_{i}\boldsymbol{k}_{j}}H_{\tau}\left(
x_{i}-x_{j}\right) \\
&  \times\text{ }\exp\left\{  -\frac{\sqrt{-1}}{2}\left(  {\sum\limits_{1\leq
i<j\leq N-1}}(\boldsymbol{k}_{i}\theta\boldsymbol{k}_{j})\mathrm{sgn}_{\tau
}(x_{i}-x_{j})\right)  \right\}  {\prod\limits_{i=2}^{N-2}}dx_{i}\text{,}%
\end{align*}
where $N\geq4$, $\boldsymbol{k}=\left(  \boldsymbol{k}_{1},\ldots
,\boldsymbol{k}_{N}\right)  $, $\boldsymbol{k}_{i}=\left(  k_{0,i}%
,\ldots,k_{l,i}\right)  $, $i=1,\ldots,N$, is the momentum vector of the
$i$-th tachyon (with Minkowski product $\boldsymbol{k}_{i}\boldsymbol{k}%
_{j}=-k_{0,i}k_{0,j}+k_{1,i}k_{1,j}+\cdots+k_{l,i}k_{l,j}$) obeying to
(\ref{momentumconservation}) and ${\prod\nolimits_{i=2}^{N-2}}dx_{i}$ is the
normalized Haar measure of $\mathbb{Q}_{p}^{N-3}$, and
\[
\mathbb{D}:=\left\{  \left(  x_{2},\ldots,x_{N-2}\right)  \in\mathbb{Q}%
_{p}^{N-3};\text{{ }}{\prod\limits_{i=2}^{N-2}}x_{i}\left(  1-x_{i}\right)
\text{ }{\prod\limits_{2\leq i<j\leq N-2}}\left(  x_{i}-x_{j}\right)  \text{
=}0\right\}  .
\]
In the bosonic string theory $l=26$, however, this choice of the dimension
does not play any role in our calculations.

In order to study the amplitude $\boldsymbol{A}^{(N)}\left(  \underline
{\boldsymbol{k}},\theta,\tau;x_{1},x_{N-1}\right)  $, we introduce
\[
\boldsymbol{s}=\left(  s_{ij}\right)  =\cup_{i=2}^{N-2}\left\{  s_{1i}%
,s_{\left(  N-1\right)  i}\right\}  \cup\cup_{2\leq i<j\leq N-2}\left\{
s_{ij}\right\}  \in\mathbb{C}^{\mathbf{d}}%
\]
a list consisting of $\mathbf{d}$ complex variables, satisfying $s_{ij}%
=s_{ji}$ for any $i$ and $j$, where
\[
\mathbf{d}:\mathbf{=}\left\{
\begin{array}
[c]{lll}%
2(N-3)+\left(
\begin{array}
[c]{c}%
N-3\\
2
\end{array}
\right)  & \text{if} & N\geq5\\
&  & \\
2 & \text{if} & N=4
\end{array}
\right.  \text{ \ \ \ \ }=\frac{N\left(  N-3\right)  }{2}.
\]
Furthermore, we introduce the variables $\widetilde{s}_{ij}\in\mathbb{R}$, for
$1\leq i<j\leq N-1$. We denote by $\widetilde{\boldsymbol{s}}=\left(
\widetilde{s}_{ij}\right)  $ for $1\leq i<j\leq N-1$. We set
\[
F(\boldsymbol{x},\boldsymbol{s},\tau):=\prod_{i=2}^{N-2}|x_{i}|_{p}^{s_{1i}%
}|1-x_{i}|_{p}^{s_{(N-1)i}}H_{\tau}(x_{i})H_{\tau}(1-x_{i})\prod_{2\leq
i<j\leq N-2}|x_{i}-x_{j}|_{p}^{s_{ij}}H_{\tau}(x_{i}-x_{j}),
\]
and
\begin{gather}
E(\boldsymbol{x},\widetilde{\boldsymbol{s}},\tau;;x_{1},x_{N-1}):=\exp\left\{
\frac{-\sqrt{-1}}{2}\left(  {\sum\limits_{2\leq j\leq N-1}}\widetilde{s}%
_{1j}\mathrm{sgn}_{\tau}(x_{1}-x_{j})\right)  \right\}  \times\nonumber\\
\exp\left\{  \frac{-\sqrt{-1}}{2}\left(  {\sum\limits_{2\leq i\leq N-2}%
}\widetilde{s}_{i\left(  N-1\right)  }\mathrm{sgn}_{\tau}(x_{i}-x_{N-1}%
)\right)  \right\}  \times\label{Formula_E}\\
\exp\left\{  \frac{-\sqrt{-1}}{2}\left(  {\sum\limits_{2\leq i<j\leq N-2}%
}\widetilde{s}_{ij}\mathrm{sgn}_{\tau}(x_{i}-x_{j})\right)  \right\}
.\nonumber
\end{gather}
Later on, we will use the convention $x_{1}=0$, $x_{N-1}=1$ and $x_{N}=\infty
$. Now, we define the Ghoshal-Kawano local zeta function as
\begin{equation}
Z^{\left(  N\right)  }(\boldsymbol{s},\widetilde{\boldsymbol{s}},\tau
;x_{1},x_{N-1})={\int\limits_{\mathbb{Q}_{p}^{N-3}\smallsetminus\mathbb{D}}%
}F(\boldsymbol{x},\boldsymbol{s},\tau)E(\boldsymbol{x},\widetilde
{\boldsymbol{s}},\tau;x_{1},x_{N-1}){\prod\limits_{i=2}^{N-2}}dx_{i}.
\label{Zeta_G_K}%
\end{equation}
For the sake of simplicity, from now on,\ we will use $\mathbb{Q}_{p}^{N-3}$
as domain of integration in (\ref{Zeta_G_K}).

By using that $\left\vert E(\boldsymbol{x},\widetilde{\boldsymbol{s}}%
,\tau;x_{1},x_{N-1})\right\vert =1$, $\left\vert H_{\tau}(x_{i})\right\vert
\leq1$, $\left\vert H_{\tau}(1-x_{i})\right\vert \leq1$, for any $i$, and that
$\left\vert H_{\tau}(x_{i}-x_{j})\right\vert \allowbreak\leq1$, for any $i$,
$j$, we have%
\begin{align*}
\left\vert Z^{\left(  N\right)  }(\boldsymbol{s},\widetilde{\boldsymbol{s}%
},\tau;x_{1},x_{N-1})\right\vert  &  \leq{\int\limits_{\mathbb{Q}_{p}^{N-3}}%
}\text{{ }}{\prod\limits_{i=2}^{N-2}}\left\vert x_{i}\right\vert
_{p}^{\operatorname{Re}(s_{1i})}\left\vert 1-x_{i}\right\vert _{p}%
^{\operatorname{Re}\left(  s_{(N-1)i}\right)  }\text{ }{\prod\limits_{2\leq
i<j\leq N-2}}\left\vert x_{i}-x_{j}\right\vert _{p}^{\operatorname{Re}\left(
s_{ij}\right)  }{\prod\limits_{i=2}^{N-2}}dx_{i}\\
&  =Z^{(N)}\left(  \operatorname{Re}(\boldsymbol{s})\right)  ,
\end{align*}
where $Z^{(N)}\left(  \boldsymbol{s}\right)  $ is the Koba-Nielsen string
amplitude studied in \cite{Bocardo-Gaspar:2016zwx}, see also
\cite{ZunigaBVeys}. Since this last integral is holomorphic in an open set
$\mathcal{K}\subset$ $\mathbb{C}^{\mathbf{d}}$, we conclude that
\[
Z^{\left(  N\right)  }(\boldsymbol{s},\widetilde{\boldsymbol{s}},\tau)\text{
is holomorphic in }\boldsymbol{s\in}\mathcal{K}\text{ for any }\widetilde
{\boldsymbol{s}}\text{, }\tau\text{, }x_{1}\text{, }x_{N-1}\text{.}%
\]
We set $T:=\left\{  2,\ldots,N-2\right\}  $, and define for $I\subseteq
T$,\textit{\ the sector attached to }$I$ as
\[
Sect(I)=\left\{  \left(  x_{2},\ldots,x_{N-2}\right)  \in\mathbb{Q}_{p}%
^{N-3};\left\vert x_{i}\right\vert _{p}\leq1\text{ }\Leftrightarrow i\in
I\right\}  .
\]
Then $\mathbb{Q}_{p}^{N-3}={\bigsqcup\nolimits_{I\subseteq T}}$ $Sect(I)$ and
\begin{equation}
Z^{\left(  N\right)  }(\boldsymbol{s},\widetilde{\boldsymbol{s}},\tau
;x_{1},x_{N-1})=\sum_{I\subseteq T}Z_{I}^{\left(  N\right)  }(\boldsymbol{s}%
,\widetilde{\boldsymbol{s}},\tau;x_{1},x_{N-1}), \label{Eq_Suma_zetas}%
\end{equation}
where
\[
Z_{I}^{\left(  N\right)  }(\boldsymbol{s},\widetilde{\boldsymbol{s}}%
,\tau;x_{1},x_{N-1}):={\int\limits_{Sect(I)}}F(\boldsymbol{x},\boldsymbol{s}%
,\tau)E(\boldsymbol{x},\widetilde{\boldsymbol{s}},\tau;x_{1},x_{N-1}%
){\prod\limits_{i=2}^{N-2}}dx_{i}.
\]
We now notice that $Z_{I}^{\left(  N\right)  }(\boldsymbol{s},\widetilde
{\boldsymbol{s}},\tau;x_{1},x_{N-1})\equiv0$ if $I\neq T$. Indeed, in the case
$I^{c}=T\smallsetminus I\neq\emptyset$, $F(\boldsymbol{x},\boldsymbol{s}%
,\tau)\equiv0$ due to the fact $H_{\tau}(x)H_{\tau}(-x)$ appears as a factor
in $F(\boldsymbol{x},\boldsymbol{s},\tau)$, and that $H_{\tau}(x)H_{\tau
}(-x)=0$. For this reason, we redefine the Ghoshal-Kawano local zeta function
as
\begin{equation}
Z^{\left(  N\right)  }(\boldsymbol{s},\widetilde{\boldsymbol{s}},\tau
;x_{1},x_{N-1})={\int\limits_{\mathbb{Z}_{p}^{N-3}}}F(\boldsymbol{x}%
,\boldsymbol{s},\tau)E(\boldsymbol{x},\widetilde{\boldsymbol{s}},\tau
;x_{1},x_{N-1}){\prod\limits_{i=2}^{N-2}}dx_{i}.
\label{Goshal-Kawano-zeta-funct}%
\end{equation}

\section{\label{Section_V}Meromorphic continuation of Ghoshal-Kawano local
zeta function}

\subsection{\label{Section_V_A}Some formulae}

For $\widetilde{s}\in\mathbb{R}$ and $x\in\mathbb{Q}_{p}\smallsetminus\left\{
0\right\}  $,
\begin{equation}
\exp\left\{  \frac{-\sqrt{-1}\widetilde{s}}{2}\mathrm{sgn}_{\tau}(x)\right\}
=\cos\left(  \frac{\widetilde{s}}{2}\right)  -\sqrt{-1}\text{ }\mathrm{sgn}%
_{\tau}(x)\sin\left(  \frac{\widetilde{s}}{2}\right)  . \label{Formula_1}%
\end{equation}
By using this formula, and the convention $x_{1}=0$, $x_{N-1}=1$, we obtain
that
\[
\exp\left\{  \frac{-\sqrt{-1}}{2}\left(
{\textstyle\sum\limits_{2\leq j\leq N-1}}
\widetilde{s}_{1j}\mathrm{sgn}_{\tau}(-x_{j})\right)  \right\}  ={\sum
\limits_{I\subseteq\left\{  2,\ldots,N-1\right\}  }}C_{I}(\widetilde
{\boldsymbol{s}}){\prod\limits_{j\in I}}\mathrm{sgn}_{\tau}(x_{j});
\]%
\[
\exp\left\{  \frac{-\sqrt{-1}}{2}\left(  {\sum\limits_{2\leq j\leq N-1}%
}\widetilde{s}_{i\left(  N-1\right)  }\mathrm{sgn}_{\tau}(x_{j}-1)\right)
\right\}  ={\sum\limits_{J\subseteq\left\{  2,\ldots,N-1\right\}  }}%
D_{J}(\widetilde{\boldsymbol{s}}){\prod\limits_{j\in J}}\mathrm{sgn}_{\tau
}(1-x_{j});
\]%
\[
\exp\left\{  \frac{-\sqrt{-1}}{2}\left(  {\sum\limits_{2\leq i<j\leq N-2}%
}\widetilde{s}_{ij}\mathrm{sgn}_{\tau}(x_{i}-x_{j})\right)  \right\}
={\sum\limits_{K\subseteq\left\{  2\leq i<j\leq N-2\right\}  }}D_{K}%
(\widetilde{\boldsymbol{s}}){\prod\limits_{i,j\in K}}\mathrm{sgn}_{\tau}%
(x_{i}-x_{j}),
\]
with the convention that ${\prod\nolimits_{j\in\emptyset}}\equiv1$. Here,
using \eqref{Formula_1}, we can see that $C_{I}$, $D_{J}$ and $D_{K}$ are
complex functions depending on $\cos\left(  \frac{\widetilde{s_{ij}}}%
{2}\right)  $ and $\sin\left(  \frac{\widetilde{s_{ij}}}{2}\right)  $. In
conclusion,
\begin{equation}
E(\boldsymbol{x},\widetilde{\boldsymbol{s}},\tau):={\sum\limits_{I,J,K}%
}E_{I,J,K}(\widetilde{\boldsymbol{s}})\text{ }{\prod\limits_{j\in I}%
}\mathrm{sgn}_{\tau}(x_{j})\text{\ }{\prod\limits_{j\in J}}\mathrm{sgn}_{\tau
}(1-x_{j}){\prod\limits_{i,j\in K}}\mathrm{sgn}_{\tau}(x_{i}-x_{j}).
\label{Formula_2}%
\end{equation}
In a similar way, we obtain that%
\begin{align}
&  \prod_{i=2}^{N-2}H_{\tau}(x_{i})H_{\tau}(1-x_{i})\prod_{2\leq i<j\leq
N-2}H_{\tau}(x_{i}-x_{j})\nonumber\\
&  ={\sum\limits_{I,J,K}}e_{I,J,K}{\prod\limits_{j\in I}}\mathrm{sgn}_{\tau
}(x_{j})\ {\prod\limits_{j\in J}}\mathrm{sgn}_{\tau}(1-x_{j})\text{{ }}%
{\prod\limits_{i,j\in K}}\mathrm{sgn}_{\tau}(x_{i}-x_{j}), \label{Formula_3}%
\end{align}
where the $e_{I,J,K}$s are constants.

\subsection{\label{Section_V_B}Meromorphic continuation of $Z^{\left(
N\right)  }(\boldsymbol{s},\widetilde{\boldsymbol{s}},\tau)$}

We assume the normalization $x_{1}=0$, $x_{N-1}=1$, $x_{N}=\infty$, and denote
the corresponding Ghoshal-Kawano zeta function as $Z^{\left(  N\right)
}(\boldsymbol{s},\widetilde{\boldsymbol{s}},\tau)$. By using formulae
(\ref{Formula_1})-(\ref{Formula_3}) and (\ref{Goshal-Kawano-zeta-funct}),
$Z^{\left(  N\right)  }(\boldsymbol{s},\widetilde{\boldsymbol{s}},\tau)$ is a
finite sum of integrals of type%

\begin{multline*}
C(\widetilde{\boldsymbol{s}})%
{\textstyle\int\limits_{\mathbb{Z}_{p}^{N-3}}}
\prod_{i=2}^{N-2}|x_{i}|_{p}^{s_{1i}}|1-x_{i}|_{p}^{s_{(N-1)i}}\prod_{2\leq
i<j\leq N-2}|x_{i}-x_{j}|_{p}^{s_{ij}}%
{\textstyle\prod\limits_{j\in I}}
\chi_{\tau}(x_{j})%
{\textstyle\prod\limits_{j\in J}}
\chi_{\tau}(1-x_{j})\\
\times%
{\textstyle\prod\limits_{i,j\in K}}
\chi_{\tau}(x_{i}-x_{j})%
{\displaystyle\prod\limits_{i=2}^{N-2}}
dx_{i},
\end{multline*}
where $C(\widetilde{\boldsymbol{s}})$ is an $\mathbb{R}$-analytic function,
$\chi_{\tau}$ denotes the trivial character or $\mathrm{sgn}_{\tau}$. This
formula implies that $Z^{\left(  N\right)  }(\boldsymbol{s},\widetilde
{\boldsymbol{s}},\tau)$ is a linear combination of multivariate Igusa local
zeta functions with coefficients in the ring of $\mathbb{R}$-analytic
functions in \ the variables $\widetilde{\boldsymbol{s}}$. Consequently, by
(\ref{Multizeta-special-case}), $Z^{\left(  N\right)  }(\boldsymbol{s}%
,\widetilde{\boldsymbol{s}},\tau)$ admits a meromorphic continuation as a
rational function in the variables $p^{-s_{1j}}$, $p^{-s_{\left(  N-1\right)
j}}$, \ $p^{-s_{ij}}$ and the real parts of its poles belong to the finite
union of hyperplanes of type
\[
\mathcal{H}=\left\{  s_{ij}\in\mathbb{C}^{\mathbf{d}};\text{ }\sum_{ij\in
M}N_{ij,k}\operatorname{Re}\left(  s_{ij}\right)  +\gamma_{k}=0,\text{ for
}k\in T\right\}  ,
\]
where $N_{ij,k}\in\mathbb{N}$, $\gamma_{k}\in\mathbb{N\smallsetminus}\left\{
0\right\}  $, and $M$, $T$ are finite sets. Furthermore, $Z^{\left(  N\right)
}(\boldsymbol{s},\widetilde{\boldsymbol{s}},\tau)$ is holomorphic in
\[%
{\textstyle\bigcap\limits_{\mathcal{H}}}
\left\{  s_{ij}\in\mathbb{C}^{\mathbf{d}};\text{ }\sum_{ij\in M}%
N_{ij,k}\operatorname{Re}\left(  s_{ij}\right)  +\gamma_{k}>0,\text{ for }k\in
T\right\}  .
\]

\subsection{\label{Section_V_C}Meromorphic continuation of $Z^{\left(
N\right)  }(\boldsymbol{s},\widetilde{\boldsymbol{s}},\tau)$ without the
normalization $x_{1}=0$, $x_{N-1}=1$}

The Ghoshal-Kawano local zeta function depends on $x_{1}$, $x_{N-1}$, i.e.
$Z^{\left(  N\right)  }(\boldsymbol{s},\widetilde{\boldsymbol{s}},\tau
;x_{1},x_{N-1})$. In \cite{Ghoshal:2004ay}, the corresponding amplitude was
considered in the case $x_{1}=0$, $x_{N-1}=1$, $x_{N}=\infty$. Our result
about the meromorphic continuation of $Z^{\left(  N\right)  }(\boldsymbol{s}%
,\widetilde{\boldsymbol{s}},\tau)$ is also valid for $Z^{\left(  N\right)
}(\boldsymbol{s},\widetilde{\boldsymbol{s}},\tau;x_{1},x_{N-1})$. Indeed, by
using that
\[
\mathbb{Z}_{p}^{N-3}=%
{\textstyle\bigsqcup\limits_{i=1}^{W}}
\boldsymbol{a}_{i}+p^{L}\mathbb{Z}_{p}^{N-3},
\]
where $W$, $L$ are positive integers, later on we will require that $L$ be
sufficiently large, and $\boldsymbol{a}_{i}\in\mathbb{Z}_{p}^{N-3}$ for any
$i$. With this notation, we have%
\[
Z^{\left(  N\right)  }(\boldsymbol{s},\widetilde{\boldsymbol{s}},\tau
;x_{1},x_{N-1})=%
{\textstyle\sum\limits_{i=1}^{W}}
Z_{\boldsymbol{a}_{i}}^{\left(  N\right)  }(\boldsymbol{s},\widetilde
{\boldsymbol{s}},\tau;x_{1},x_{N-1}),
\]
where
\[
Z_{\boldsymbol{b}}^{\left(  N\right)  }(\boldsymbol{s},\widetilde
{\boldsymbol{s}},\tau;x_{1},x_{N-1}):=%
{\textstyle\int\limits_{\boldsymbol{b}+p^{L}\mathbb{Z}_{p}^{N-3}}}
F(\boldsymbol{x},\boldsymbol{s},\tau)E(\boldsymbol{x},\widetilde
{\boldsymbol{s}},\tau;x_{1},x_{N-1})%
{\displaystyle\prod\limits_{i=2}^{N-2}}
dx_{i},
\]
see (\ref{Formula_E}). The meromorphic continuation of $\ $ $Z_{\boldsymbol{b}%
}^{\left(  N\right)  }(\boldsymbol{s},\widetilde{\boldsymbol{s}},\tau
;x_{1},x_{N-1})$ can be obtained by the methods presented in Sections
(\ref{Section_V_A})-(\ref{Section_V_B}),\ by computing a Taylor expansion of
the polynomial $\prod_{i=2}^{N-2}x_{i}\left(  1-x_{i}\right)  \prod_{2\leq
i<j\leq N-2}\left(  x_{i}-x_{j}\right)  $ near $\boldsymbol{b}$.

\section{\label{Section_VI}Explicit computation of $Z^{\left(  4\right)
}(\boldsymbol{s},\widetilde{\boldsymbol{s}},\tau,)$}

In this section we compute the Ghoshal-Kawano local zeta function for four
points:%
\[
Z^{\left(  4\right)  }(\boldsymbol{s},\widetilde{\boldsymbol{s}},\tau
)=\exp\big\{\frac{i}{2}\widetilde{s}_{13}\big\}%
{\textstyle\int\limits_{\mathbb{Z}_{p}}}
\left\vert x_{2}\right\vert _{p}^{s_{12}}\left\vert 1-x_{2}\right\vert
_{p}^{s_{32}}H_{\tau}\left(  x_{2}\right)  H_{\tau}\left(  1-x_{2}\right)
E^{\left(  4\right)  }\left(  x_{2},\widetilde{\boldsymbol{s}},\tau\right)
dx_{2},
\]
where%
\[
E^{\left(  4\right)  }\left(  x_{2},\widetilde{\boldsymbol{s}},\tau\right)
:=E^{\left(  4\right)  }\left(  x_{2},\widetilde{s}_{12},\widetilde{s}%
_{32},\tau\right)  =\exp\bigg\{\frac{i}{2}\bigg(\widetilde{s}_{12}%
\mathrm{sgn}_{\tau}(x_{2})+\widetilde{s}_{23}\mathrm{sgn}_{\tau}%
(1-x_{2})\bigg)\bigg\}.
\]
We recall that Ghoshal and Kawano take $x_{1}=0$, $x_{3}=1$, $x_{4}=\infty$.
By using the fact that $\mathrm{sgn}_{\tau}(y)\in\left\{  1,-1\right\}  $ and
$H_{\tau}\left(  y\right)  \in\left\{  0,1\right\}  $, one verifies that%
\begin{align*}
\exp\bigg\{\frac{i}{2}\bigg(\widetilde{s}_{12}\mathrm{sgn}_{\tau}%
(x_{2})\bigg)\bigg\}H_{\tau}\left(  x_{2}\right)   &  =\exp\bigg(\frac{i}%
{2}\widetilde{s}_{12}\bigg)H_{\tau}\left(  x_{2}\right)  ,\\
\exp\bigg\{\frac{i}{2}\bigg(\widetilde{s}_{23}\mathrm{sgn}_{\tau}%
(1-x_{2})\bigg)\bigg\}H_{\tau}\left(  1-x_{2}\right)   &  =\exp\left(
\frac{i}{2}\widetilde{s}_{23}\right)  H_{\tau}\left(  1-x_{2}\right)  ,
\end{align*}
and consequently,%
\[
E^{\left(  4\right)  }\left(  \widetilde{s}_{12},\widetilde{s}_{32}\right)
=\exp\left\{  \frac{i}{2}\bigg(\widetilde{s}_{12}+\widetilde{s}_{23}%
\bigg)\right\}  ,
\]
and%
\[
Z^{\left(  4\right)  }(\boldsymbol{s},\widetilde{\boldsymbol{s}},\tau
)=\exp\left\{  \frac{i}{2}\bigg(\widetilde{s}_{13}+\widetilde{s}%
_{12}+\widetilde{s}_{23}\bigg)\right\}
{\textstyle\int\limits_{\mathbb{Z}_{p}}}
\left\vert x_{2}\right\vert _{p}^{s_{12}}\left\vert 1-x_{2}\right\vert
_{p}^{s_{32}}H_{\tau}\left(  x_{2}\right)  H_{\tau}\left(  1-x_{2}\right)
dx_{2}.
\]
We first compute some $p$-adic integrals needed in this section.

\subsection{\label{Section_VI_A}Some $p$-adic integrals}

\subsubsection{Formula 1}

Assume that $S\subset Z_{p}\smallsetminus\left\{  0\right\}  $ satisfies
$-S=S$. Then, for $\tau\in\left\{  p,\varepsilon p\right\}  $%
\[%
{\textstyle\int\limits_{S}}
\left\vert x_{2}\right\vert _{p}^{s_{12}}\mathrm{sgn}_{\tau}(x_{2})dx_{2}=0.
\]
This formula follows from changing variables as $x_{2}=-y$ and using the fact
that \textrm{sgn}$_{\tau}(-y)=-\mathrm{sgn}_{\tau}(y)$.

\subsubsection{Formula 2}

If $p\equiv3\operatorname{mod}4$ and $\tau\in\left\{  p,\varepsilon p\right\}
$, then
\[
S(\tau,p):=\frac{1}{p}\sum\limits_{j=2}^{p-1}H_{\tau}(j)H_{\tau}(1-j)=
\frac{p-3}{4p}
\]
From table (\ref{Table}), for $j=2,\ldots,p-1$,
\[
H_{\tau}(j)H_{\tau}(1-j)= \frac{1}{4}\left\{  1+\left(  \frac{j}{p}\right)
\right\}  \left\{  1-\left(  \frac{j-1}{p}\right)  \right\}
\]
and thus
\[
S(\tau,p):=\frac{1}{4p}\left\{  p-2+\sum\limits_{j=2}^{p-1}\left(  \frac{j}%
{p}\right)  -\sum\limits_{j=2}^{p-1}\left(  \frac{j-1}{p}\right)
-\sum\limits_{j=2}^{p-1}\left(  \frac{j}{p}\right)  \left(  \frac{j-1}%
{p}\right)  \right\}  .
\]
Now by using that $\sum\limits_{k=1}^{p-1}\left(  \frac{k}{p}\right)  =0$, we
get that%
\[
\sum\limits_{j=2}^{p-1}\left(  \frac{j}{p}\right)  =-1\text{ and }%
\sum\limits_{j=2}^{p-1}\left(  \frac{j-1}{p}\right)  =\sum\limits_{k=1}%
^{p-2}\left(  \frac{k}{p}\right)  =-\left(  \frac{p-1}{p}\right)  =1,
\]
and thus%
\[
S(\tau,p)=\frac{1}{4p}\left\{  p-4-\sum\limits_{k=1}^{p-2}\left(  \frac
{k+1}{p}\right)  \left(  \frac{k}{p}\right)  \right\}  .
\]
To compute
\[
L(\tau,p):=\sum\limits_{k=1}^{p-2}\left(  \frac{k+1}{p}\right)  \left(
\frac{k}{p}\right)  ,
\]
we define%
\[
A_{ij}=\left\{  a\in\left\{  1,\ldots,p-2\right\}  ;\left(  \frac{a}%
{p}\right)  =\left(  -1\right)  ^{i}\text{ and }\left(  \frac{a+1}{p}\right)
=\left(  -1\right)  ^{j}\right\}  ,
\]
then $\left\{  1,\ldots,p-2\right\}  =A_{00}%
{\textstyle\bigsqcup}
A_{01}%
{\textstyle\bigsqcup}
A_{10}%
{\textstyle\bigsqcup}
A_{11}$ and
\[
L(\tau,p)=\#A_{00}-\#A_{01}-\#A_{10}+\#A_{11}.
\]
Now, if $p\equiv3\operatorname{mod}4$, then%
\begin{equation}
\#A_{00}=\#A_{10}=\#A_{11}=\frac{p-3}{4}\text{, and }\#A_{01}=\frac{p+1}{4},
\label{Formula_A_s}%
\end{equation}
see e.g. \cite[ Chapter 9, Exercise 5 in p. 201 ]{NumberTheory}, and
therefore
\[
L(\tau,p)=\#A_{00}-\#A_{01}=-1\text{, and }S(\tau,p)=\frac{1}{4p}(p-3).
\]

\subsubsection{Formula 3}

\label{Formula3}

Set
\[
I(\boldsymbol{s},\tau)=%
{\textstyle\int\limits_{\mathbb{Z}_{p}}}
\left\vert x_{2}\right\vert _{p}^{s_{12}}\left\vert 1-x_{2}\right\vert
_{p}^{s_{32}}H_{\tau}\left(  x_{2}\right)  H_{\tau}\left(  1-x_{2}\right)
dx_{2}.
\]
Then%
\begin{equation}
I(\boldsymbol{s},\tau)= \frac{p-3}{4p}+\frac{p^{-1-s_{12}}\left(
1-p^{-1}\right)  }{2\left(  1-p^{-1-s_{12}}\right)  }+\frac{p^{-1-s_{32}%
}\left(  1-p^{-1}\right)  }{2\left(  1-p^{-1-s_{32}}\right)  }
\label{Formula_I}%
\end{equation}
By using the partition%
\begin{equation}
\mathbb{Z}_{p}=%
{\textstyle\bigsqcup\nolimits_{j=0}^{p-1}}
j+p\mathbb{Z}_{p} \label{Partition}%
\end{equation}
and the fact that%
\begin{equation}
H_{\tau}\left(  x_{2}\right)  \mid_{j+p\mathbb{Z}_{p}}=H_{\tau}\left(
j\right)  \text{ for }j\neq0\text{ \ and }H_{\tau}\left(  1-x_{2}\right)
\mid_{j+p\mathbb{Z}_{p}}=H_{\tau}\left(  1-j\right)  \text{ for }%
j\neq1\text{,} \label{Formula_H_j}%
\end{equation}
we have
\begin{equation}
I(\boldsymbol{s},\tau)=%
{\textstyle\sum\nolimits_{j=0}^{p-1}}
I_{j}(\boldsymbol{s},\tau), \label{Formula_I_suma}%
\end{equation}
where
\[
I_{j}(\boldsymbol{s},\tau)=%
{\textstyle\int\limits_{j+p\mathbb{Z}_{p}}}
\left\vert x_{2}\right\vert _{p}^{s_{12}}\left\vert 1-x_{2}\right\vert
_{p}^{s_{32}}H_{\tau}\left(  x_{2}\right)  H_{\tau}\left(  1-x_{2}\right)
dx_{2}.
\]
If $j\neq0,1$, then
\begin{equation}
I_{j}(\boldsymbol{s},\tau)=p^{-1}H_{\tau}\left(  j\right)  H_{\tau}\left(
1-j\right)  . \label{Formula_I_j}%
\end{equation}
If $j=0$, then by using Formula 1,%
\begin{align}
I_{0}(\boldsymbol{s},\tau)  &  =%
{\textstyle\int\limits_{p\mathbb{Z}_{p}}}
\left\vert x_{2}\right\vert _{p}^{s_{12}}H_{\tau}\left(  x_{2}\right)
dx_{2}=\frac{1}{2}%
{\textstyle\int\limits_{p\mathbb{Z}_{p}}}
\left\vert x_{2}\right\vert _{p}^{s_{12}}dx_{2}+\frac{1}{2}%
{\textstyle\int\limits_{p\mathbb{Z}_{p}}}
\left\vert x_{2}\right\vert _{p}^{s_{12}}\mathrm{sgn}_{\tau}(x_{2}%
)dx_{2}\nonumber\\
&  =\frac{1}{2}%
{\textstyle\int\limits_{p\mathbb{Z}_{p}}}
\left\vert x_{2}\right\vert _{p}^{s_{12}}dx_{2}=\frac{1}{2}\frac{p^{-1-s_{12}%
}\left(  1-p^{-1}\right)  }{1-p^{-1-s_{12}}}. \label{Formula_I_0}%
\end{align}
The case $j=1$ is similar to the case $j=0$,%
\begin{equation}
I_{1}(\boldsymbol{s},\tau)=%
{\textstyle\int\limits_{1+p\mathbb{Z}_{p}}}
\left\vert 1-x_{2}\right\vert _{p}^{s_{32}}H_{\tau}\left(  1-x_{2}\right)
dx_{2}=\frac{1}{2}\frac{p^{-1-s_{32}}\left(  1-p^{-1}\right)  }{1-p^{-1-s_{32}%
}}. \label{Formula_I_1}%
\end{equation}
Formula (\ref{Formula_I}) follows from (\ref{Formula_I_suma}) by using
(\ref{Formula_I_j})-(\ref{Formula_I_1}) and Formula 2.

\subsection{\label{Section_VI_B}Computation of $Z^{\left(  4\right)
}(\boldsymbol{s},\widetilde{\boldsymbol{s}},\tau)$}

In conclusion,%
\begin{align}
Z^{\left(  4\right)  }(\boldsymbol{s},\widetilde{\boldsymbol{s}},\tau)  &
=\exp\left\{  \frac{i}{2}\big(\widetilde{s}_{13}+\widetilde{s}_{12}%
+\widetilde{s}_{23}\big)\right\}  I(\boldsymbol{s},\tau)\nonumber\\
&  =\exp\left\{  \frac{i}{2}\big(\widetilde{s}_{13}+\widetilde{s}%
_{12}+\widetilde{s}_{23}\big)\right\}  \left(  \frac{p-3}{4p}+\frac
{p^{-1-s_{12}}\left(  1-p^{-1}\right)  }{2\left(  1-p^{-1-s_{12}}\right)
}+\frac{p^{-1-s_{32}}\left(  1-p^{-1}\right)  }{2\left(  1-p^{-1-s_{32}%
}\right)  }\right)  \label{cuatropuntos}%
\end{align}
is holomorphic in
\begin{equation}
\operatorname{Re}(s_{12})>-1\text{ and }\operatorname{Re}(s_{32})>-1.
\label{Band}%
\end{equation}
The above formula for $Z^{\left(  4\right)  }(\boldsymbol{s},\widetilde
{\boldsymbol{s}},\tau)$ was also obtained in \cite{Ghoshal:2004ay}.

\section{\label{Section_VII}Explicit computation of $Z^{(5)}(\boldsymbol{s}%
,\widetilde{\boldsymbol{s}},\tau)$}

In \ this section, using the normalization $x_{1}=0$, $x_{4}=1$ and
$x_{5}=\infty$, we compute the amplitude for five points:
\[
Z^{(5)}(\boldsymbol{s},\widetilde{\boldsymbol{s}},\tau)=\int_{\mathbb{Z}%
_{p}^{2}}E^{(5)}(x_{2},x_{3},\widetilde{\boldsymbol{s}},\tau)F^{(5)}%
(x_{2},x_{3},\boldsymbol{s},\tau)dx_{2}dx_{3},
\]
with
\begin{align*}
E^{(5)}(x_{2},x_{3},\widetilde{\boldsymbol{s}},\tau)  &  =\exp\left\{
\frac{-\sqrt{-1}}{2}\bigg(\widetilde{s}_{14}\mathrm{sgn}_{\tau}(-1)+\widetilde
{s}_{12}\mathrm{sgn}_{\tau}(-x_{2})+\widetilde{s}_{13}\mathrm{sgn}_{\tau
}(-x_{3})\bigg)\right\} \\
&  \times\exp\left\{  \frac{-\sqrt{-1}}{2}\bigg(\widetilde{s}_{42}%
\mathrm{sgn}_{\tau}(1-x_{2})+\tilde{s}_{43}\mathrm{sgn}_{\tau}(1-x_{3}%
)+\tilde{s}_{23}\mathrm{sgn}_{\tau}(x_{2}-x_{3})\bigg)\right\}
\end{align*}
and
\begin{align*}
F^{(5)}(x_{2},x_{3},\boldsymbol{s},\tau)  &  =|x_{2}|_{p}^{s_{12}}|x_{3}%
|_{p}^{s_{13}}|1-x_{2}|_{p}^{s_{42}}|1-x_{3}|_{p}^{s_{43}}|x_{2}-x_{3}%
|_{p}^{s_{23}}\\
&  \times H_{\tau}(x_{2})H_{\tau}(x_{3})H_{\tau}(1-x_{2})H_{\tau}%
(1-x_{3})H_{\tau}(x_{2}-x_{3}).
\end{align*}
Using the reasoning given at the beginning of the previous section we have
\[
E^{(5)}(\widetilde{\boldsymbol{s}})=\exp\left\{  \frac{\sqrt{-1}}%
{2}\bigg(\widetilde{s}_{14}+\widetilde{s}_{12}+\widetilde{s}_{13}%
+\widetilde{s}_{24}+\tilde{s}_{34}+\tilde{s}_{32}\bigg)\right\}
\]
and then
\begin{equation}
Z^{(5)}(\boldsymbol{s},\widetilde{\boldsymbol{s}},\tau)=E^{(5)}(\widetilde
{\boldsymbol{s}})L(\boldsymbol{s},\tau), \label{Zeta_5}%
\end{equation}
where
\begin{equation}
L(\boldsymbol{s},\tau)=\int_{\mathbb{Z}_{p}^{2}}F^{(5)}(x_{2},x_{3}%
,\boldsymbol{s},\tau)dx_{2}dx_{3}. \label{Zeta_5A}%
\end{equation}
First we give some formulae needed in the following calculations.

\subsection{\label{Section_VII_A}More $p$-adic Sums and Integrals}

\subsubsection{Formula 4}

For $A\subset\{1,2,\dots,p-1\}$, by using that $H_{\tau}(x)=\frac{1}%
{2}(1+\mathrm{sgn}_{\tau}(x))$, \ where the sign function $\mathrm{sgn}_{\tau
}$ is give in Table (\ref{Table}), \ and that $\mathrm{sgn}_{\tau
}(-x)=-\mathrm{sgn}_{\tau}(x)$, for $p\equiv3$ mod $4$ and $\tau
\in\{p,\varepsilon p\}$, we have%
\[
V(A,p,\tau):=\sum_{\substack{i,j\in A\\i\neq j}}^{p-1}H_{\tau}(i-j)=\frac
{(\#A)(\#A-1)}{2}=\binom{\#A}{2}.
\]
Indeed
\begin{align*}
V(A,p,\tau)  &  =\sum_{\substack{i,j\in A\\j<i}}H_{\tau}(i-j)+\sum
_{\substack{i,j\in A\\i<j}}H_{\tau}(-(j-i))=\sum_{\substack{i,j\in
A\\j<i}}\left[  H_{\tau}(i-j)+H_{\tau}(-(i-j))\right] \\
&  =\frac{1}{2}\sum_{\substack{i,j\in A\\j<i}}\left[  1+\left(  \frac{i-j}%
{p}\right)  +1-\left(  \frac{i-j}{p}\right)  \right]  =\sum_{\substack{i,j\in
A\\j<i}}1=\binom{\#A}{2}.
\end{align*}

\subsubsection{Formula 5}

If $p\equiv3$ mod $4$ and $\tau\in\{p,\varepsilon p\}$, then
\begin{align*}
T(p,\tau)  &  :=\frac{1}{p^{2}}\sum_{\substack{i,j=2\\i\neq j}}^{p-1}H_{\tau
}(i)H_{\tau}(1-i)H_{\tau}(j)H_{\tau}(1-j)H_{\tau}(i-j)\\
&  =\frac{(p-3)(p-7)}{32p^{2}}.
\end{align*}
We define $B:=\left\{  k\in\left\{  2,3,\dots,p-1\right\}  ;H_{\tau}%
(k)H_{\tau}(1-k)=1\right\}  $. Then, by using the results and notation given
in the proof of Formula 2, we have $\#B=\#A_{10}=\frac{p-3}{4}$, and
\[
T(p,\tau)=\frac{1}{p^{2}}\sum_{\substack{i,j\in B\\i\neq j}}H_{\tau
}(i-j)=\frac{1}{p^{2}}\binom{\#B}{2}=\frac{1}{2p^{2}}\left(  \frac{p-3}%
{4}\right)  \left(  \frac{p-7}{4}\right)  =\frac{\left(  p-3\right)  \left(
p-7\right)  }{32p^{2}}.
\]

\subsubsection{Formula 6}

We set for $a$, $b$, $c\in\mathbb{C}$,%
\[
L_{00}(a,b,c):=\frac{1}{8}\int_{\left(  p\mathbb{Z}_{p}\right)  ^{2}}%
|x_{2}|_{p}^{a}|x_{3}|_{p}^{b}|x_{2}-x_{3}|_{p}^{c}dx_{2}dx_{3}\text{, for
}\operatorname{Re}(a)\text{, }\operatorname{Re}(b)\text{, }\operatorname{Re}%
(c)>0\text{.}%
\]
Then $L_{00}(a,b,c)$ has a meromorphic continuation to the whole complex plane
given by%
\begin{align*}
L_{00}(a,b,c)  &  =\frac{1}{8}\frac{p^{-a-b-c-2}\left(  1-p^{-1}\right)
}{1-p^{-a-b-c-2}}\left\{  p^{-1}\left(  p-2\right)  +\frac{p^{-1-a}\left(
1-p^{-1}\right)  }{1-p^{-1-a}}+\frac{p^{-1-b}\left(  1-p^{-1}\right)
}{1-p^{-1-b}}\right. \\
&  \left.  +\frac{p^{-1-c}\left(  1-p^{-1}\right)  }{1-p^{-1-c}}\right\}  .
\end{align*}
In order to compute $L_{00}(a,b,c)$, we introduce the following subsets:%
\[
A:=\left\{  \left(  x_{2},x_{3}\right)  \in\left(  p\mathbb{Z}_{p}\right)
^{2};\left\vert \frac{x_{2}}{x_{3}}\right\vert _{p}\leq1\right\}  ,
\]%
\[
B:=\left\{  \left(  x_{2},x_{3}\right)  \in\left(  p\mathbb{Z}_{p}\right)
^{2};\left\vert \frac{x_{3}}{x_{2}}\right\vert _{p}<1\right\}  .
\]
Then
\begin{equation}
\left(  p\mathbb{Z}_{p}\right)  ^{2}\smallsetminus\left\{  \left(  x_{2}%
,x_{3}\right)  \in\left(  p\mathbb{Z}_{p}\right)  ^{2};x_{2}x_{3}=0\right\}
=A%
{\textstyle\bigsqcup}
B, \label{Partition_2}%
\end{equation}
and $L_{00}(a,b,c)=L_{00}^{\left(  A\right)  }(a,b,c)+L_{00}^{\left(
B\right)  }(a,b,c)$, where%
\[
L_{00}^{\left(  A\right)  }(a,b,c):=\frac{1}{8}\int_{A}|x_{2}|_{p}^{a}%
|x_{3}|_{p}^{b}|x_{2}-x_{3}|_{p}^{c}dx_{2}dx_{3}\text{, \ }%
\]
and
\[
L_{00}^{\left(  B\right)  }(a,b,c):=\frac{1}{8}\int_{B}|x_{2}|_{p}^{a}%
|x_{3}|_{p}^{b}|x_{2}-x_{3}|_{p}^{c}dx_{2}dx_{3}.
\]
We compute first $L_{00}^{\left(  A\right)  }(a,b,c)$, by using the following
change of variables:%
\begin{equation}
\text{ }x_{2}=uv,\text{ }x_{3}=u\text{.} \label{Change_var_1}%
\end{equation}
Then $dx_{2}dx_{3}=\left\vert u\right\vert _{p}dudv$ and
\begin{align}
L_{00}^{\left(  A\right)  }(a,b,c)  &  =\frac{1}{8}\int_{p\mathbb{Z}_{p}%
\times\mathbb{Z}_{p}}|u|_{p}^{a+b+c+1}|v|_{p}^{a}|v-1|_{p}^{c}dudv\nonumber\\
&  =\frac{1}{8}\left\{  \int_{p\mathbb{Z}_{p}}|u|_{p}^{a+b+c+1}du\right\}
\left\{  \int_{\mathbb{Z}_{p}}|v|_{p}^{a}|v-1|_{p}^{c}dv\right\} \nonumber\\
&  =:\frac{1}{8}\frac{p^{-a-b-c-2}\left(  1-p^{-1}\right)  }{1-p^{-a-b-c-2}%
}J\left(  a,c\right)  . \label{Formula_L00_A}%
\end{align}
By using partition (\ref{Partition}),%
\[
J\left(  a,c\right)  =%
{\textstyle\sum\limits_{i=0}^{p-1}}
J_{i}\left(  a,c\right)  .
\]
For $i\neq0,1$,
\[
J_{i}\left(  a,c\right)  =\int_{i+p\mathbb{Z}_{p}}|v|_{p}^{a}|v-1|_{p}%
^{c}dv=p^{-1},
\]
and the contribution of all these integrals is \
\begin{equation}%
{\textstyle\sum\limits_{i=2}^{p-1}}
J_{i}\left(  a,c\right)  =p^{-1}\left(  p-2\right)  . \label{Formula_L00_A_1}%
\end{equation}
For $i=0$,
\begin{equation}
J_{0}\left(  a,c\right)  =\int_{p\mathbb{Z}_{p}}|v|_{p}^{a}dv=\frac
{p^{-1-a}\left(  1-p^{-1}\right)  }{1-p^{-1-a}}. \label{Formula_L00_A_2}%
\end{equation}
For $i=1$,
\begin{equation}
J_{1}\left(  a,c\right)  =\int_{1+p\mathbb{Z}_{p}}|v-1|_{p}^{c}dv=\frac
{p^{-1-c}\left(  1-p^{-1}\right)  }{1-p^{-1-c}}. \label{Formula_L00_A_3}%
\end{equation}
Therefore, from (\ref{Formula_L00_A})-(\ref{Formula_L00_A_3}),%
\[
L_{00}^{\left(  A\right)  }(a,b,c)=\frac{1}{8}\frac{p^{-a-b-c-2}\left(
1-p^{-1}\right)  }{1-p^{-a-b-c-2}}\left\{  p^{-1}\left(  p-2\right)
+\frac{p^{-1-a}\left(  1-p^{-1}\right)  }{1-p^{-1-a}}+\frac{p^{-1-c}\left(
1-p^{-1}\right)  }{1-p^{-1-c}}\right\}  .
\]
Now we compute $L_{00}^{\left(  B\right)  }(a,b,c)$, by using the following
change of variables:%
\begin{equation}
x_{2}=t,\text{ \ }x_{3}=zt. \label{Change_var_2}%
\end{equation}
Then $dx_{2}dx_{3}=\left\vert t\right\vert _{p}dzdt$ and
\begin{align*}
L_{00}^{\left(  B\right)  }(a,b,c)  &  =\frac{1}{8}\int_{\left(
p\mathbb{Z}_{p}\right)  ^{2}}|t|_{p}^{a+b+c+1}|z|_{p}^{b}|1-z|_{p}%
^{c}dzdt=\frac{1}{8}\int_{\left(  p\mathbb{Z}_{p}\right)  ^{2}}|t|_{p}%
^{a+b+c+1}|z|_{p}^{b}|dzdt\\
&  =\frac{1}{8}\frac{p^{-a-b-c-2}\left(  1-p^{-1}\right)  }{1-p^{-a-b-c-2}%
}\frac{p^{-1-b}\left(  1-p^{-1}\right)  }{1-p^{-1-b}}.
\end{align*}

\subsubsection{Formula 7}

For $a$, $b$, $c\in\mathbb{C}$,%
\begin{align*}
L_{00}^{\left(  1\right)  }(a,b,c,\tau)  &  :=\frac{1}{8}\int_{\left(
p\mathbb{Z}_{p}\right)  ^{2}}|x_{2}|_{p}^{a}|x_{3}|_{p}^{b}|x_{2}-x_{3}%
|_{p}^{c}\mathrm{sgn}_{\tau}(x_{2})\mathrm{sgn}_{\tau}(x_{3})dx_{2}dx_{3}\\
&  =\frac{1}{8}\frac{p^{-a-b-c-2}\left(  1-p^{-1}\right)  }{1-p^{-a-b-c-2}%
}\left\{  -p^{-1}+\frac{p^{-1-c}\left(  1-p^{-1}\right)  }{1-p^{-1-c}%
}\right\}  \text{, }%
\end{align*}
for $\operatorname{Re}(a)$, $\operatorname{Re}(b)$, $\operatorname{Re}(c)>0$.
By using partition (\ref{Partition_2}), we get that \ $L_{00}^{\left(
1\right)  }(a,b,c,\tau)=L_{00}^{\left(  1,A\right)  }(a,b,c,\tau
)+L_{00}^{\left(  1,B\right)  }(a,b,c,\tau)$. We compute integral
$L_{00}^{\left(  1,A\right)  }(a,b,c,\tau)$, respectively $L_{00}^{\left(
1,B\right)  }(a,b,c,\tau)$, by using change of variables (\ref{Change_var_1}),
respectively (\ref{Change_var_2}), as follows:%
\begin{align*}
L_{00}^{\left(  1,A\right)  }(a,b,c,\tau)  &  =\frac{1}{8}\left\{
\int_{p\mathbb{Z}_{p}}|u|_{p}^{a+b+c+1}du\right\}  \left\{  \int
_{\mathbb{Z}_{p}}|v|_{p}^{a}|v-1|_{p}^{c}\mathrm{sgn}_{\tau}(v)dv\right\} \\
&  =\frac{1}{8}\frac{p^{-a-b-c-2}\left(  1-p^{-1}\right)  }{1-p^{-a-b-c-2}%
}K(a,c,\tau).
\end{align*}
By using partition (\ref{Partition}),%
\[
K(a,c,\tau)=%
{\textstyle\sum\limits_{j=0}^{p-1}}
K_{j}(a,c,\tau).
\]
For $j\neq0,1$, $K_{j}(a,c,\tau)=p^{-1}\mathrm{sgn}_{\tau}(j)$, thus, the
contribution of all these integrals is%
\[%
{\textstyle\sum\limits_{j=2}^{p-1}}
K_{j}(a,c,\tau)=p^{-1}%
{\textstyle\sum\limits_{j=2}^{p-1}}
\mathrm{sgn}_{\tau}(j)=p^{-1}%
{\textstyle\sum\limits_{j=2}^{p-1}}
(\frac{j}{p})=-p^{-1}.
\]
For $j=0$, by using the Formula $1$, $K_{0}(a,c,\tau)=0$. For $j=1$,
\begin{align*}
K_{1}(a,c,\tau)  &  =\int_{1+p\mathbb{Z}_{p}}|v-1|_{p}^{c}\mathrm{sgn}_{\tau
}(v)dv=\int_{1+p\mathbb{Z}_{p}}|v-1|_{p}^{c}dv\\
&  =\int_{p\mathbb{Z}_{p}}|v|_{p}^{c}dv=\frac{p^{-1-c}\left(  1-p^{-1}\right)
}{1-p^{-1-c}}.
\end{align*}
In conclusion,%
\[
L_{00}^{\left(  1,A\right)  }(a,b,c,\tau)=\frac{1}{8}\frac{p^{-a-b-c-2}\left(
1-p^{-1}\right)  }{1-p^{-a-b-c-2}}\left\{  -p^{-1}+\frac{p^{-1-c}\left(
1-p^{-1}\right)  }{1-p^{-1-c}}\right\}  .
\]
Now, by Formula $1$,%
\[
L_{00}^{\left(  1,B\right)  }(a,b,c,\tau)=\left\{  \frac{1}{8}\int
_{p\mathbb{Z}_{p}}|t|_{p}^{a+b+c+1}dt\right\}  \left\{  \int_{p\mathbb{Z}_{p}%
}|z|_{p}^{b}\text{ }\mathrm{sgn}_{\tau}(z)dz\right\}  =0.
\]

\subsubsection{Formula 8}

For $a$, $b$, $c\in\mathbb{C}$, we set%
\[
L_{00}^{\left(  2\right)  }(a,b,c,\tau):=\frac{1}{8}\int_{\left(
p\mathbb{Z}_{p}\right)  ^{2}}|x_{2}|_{p}^{a}|x_{3}|_{p}^{b}|x_{2}-x_{3}%
|_{p}^{c}\text{ }\mathrm{sgn}_{\tau}(x_{2})\mathrm{sgn}_{\tau}(x_{2}%
-x_{3})dx_{2}dx_{3}.
\]
Then
\[
L_{00}^{\left(  2\right)  }(a,b,c,\tau)=L_{00}^{\left(  1\right)  }%
(a,c,b,\tau).
\]
This identity is obtained by changing variables as $u=x_{2}$, $v=x_{2}-x_{3}%
$,\ and using Formula 7.

\subsubsection{Formula 9}

For $a$, $b$, $c\in\mathbb{C}$, we set
\[
L_{00}^{\left(  3\right)  }(a,b,c,\tau):=\frac{1}{8}\int_{\left(
p\mathbb{Z}_{p}\right)  ^{2}}|x_{2}|_{p}^{a}|x_{3}|_{p}^{b}|x_{2}-x_{3}%
|_{p}^{c}\text{ }\mathrm{sgn}_{\tau}(x_{3})\mathrm{sgn}_{\tau}(x_{2}%
-x_{3})dx_{2}dx_{3}.
\]
Then $\ $%
\[
L_{00}^{\left(  3\right)  }(a,b,c,\tau)=-L_{00}^{\left(  2\right)
}(b,a,c,\tau).
\]
This formula follows from Formula 8 by changing variables as $\left(
x_{2},x_{3}\right)  \rightarrow\left(  x_{3},x_{2}\right)  $.

\subsection{\label{Section_VII_B}Computation of $Z^{(5)}(\boldsymbol{s}%
,\widetilde{\boldsymbol{s}},\tau)$}

The \ computation of $Z^{(5)}(\boldsymbol{s},\widetilde{\boldsymbol{s}},\tau)$
is reduced to the computation of integral $L(\boldsymbol{s},\tau)$, see
(\ref{Zeta_5})-(\ref{Zeta_5A}). By using the partition
\[
\mathbb{Z}_{p}^{2}=\bigsqcup_{i,j=0}^{p-1}\left(  i+p\mathbb{Z}_{p}\right)
\times\left(  j+p\mathbb{Z}_{p}\right)  ,
\]
we have
\[
L(\boldsymbol{s},\tau)=\sum_{i,j=0}^{p-1}L_{ij}(\boldsymbol{s},\tau)
\]
where
\[
L_{ij}(\boldsymbol{s},\tau)=\int_{i+p\mathbb{Z}_{p}\times j+p\mathbb{Z}_{p}%
}F^{(5)}(x_{2},x_{3},\boldsymbol{s},\tau)dx_{2}dx_{3}.
\]
The calculation of these integrals is achieved by considering several cases.

\textbf{Case }$i,j\in\left\{  2,3,\ldots,p-1\right\}  $\textbf{ and }$i\neq j$.

In this case, \ by using that $H_{\tau}\mid_{i+p\mathbb{Z}_{p}}=H_{\tau}(i)$
for $i\in\left\{  1,\ldots,p-1\right\}  $,
\[
L_{ij}(\boldsymbol{s},\tau)=p^{-2}H_{\tau}(i)H_{\tau}(1-i)H_{\tau}(j)H_{\tau
}(1-i)H_{\tau}(i-j).
\]
Now by using Formula $5$, the contribution of all these integrals is%
\begin{equation}
\sum_{\substack{i,j=2\\i\neq j}}^{p-1}L_{ij}(\boldsymbol{s},\tau
)=\frac{(p-3)(p-7)}{32p^{2}}. \label{Formula_L_ij}%
\end{equation}

\textbf{Case }$i,j\in\left\{  2,3,\ldots,p-1\right\}  $\textbf{ and }$i=j$.

In this case, by using (\ref{Formula_I_0}),
\begin{align*}
L_{ii}(\boldsymbol{s},\tau)  &  =H_{\tau}^{2}(i)H_{\tau}^{2}(1-i)\int
_{i+p\mathbb{Z}_{p}\times i+p\mathbb{Z}_{p}}|x_{2}-x_{3}|_{p}^{s_{23}}H_{\tau
}(x_{2}-x_{3})dx_{2}dx_{3}\\
&  =H_{\tau}(i)H_{\tau}(1-i)\int_{\left(  p\mathbb{Z}_{p}\right)  ^{2}}%
|x_{2}-x_{3}|_{p}^{s_{23}}H_{\tau}(x_{2}-x_{3})dx_{2}dx_{3}\\
&  =p^{-1}H_{\tau}(i)H_{\tau}(1-i)\int_{ p\mathbb{Z}_{p}}|x_{2}|_{p}^{s_{23}%
}H_{\tau}(x_{2})dx_{2}=p^{-1}H_{\tau}(i)H_{\tau}(1-i)I_{0}(s_{23},\tau)\\
&  =p^{-1}H_{\tau}(i)H_{\tau}(1-i)\frac{p^{-1-s_{23}}(1-p^{-1})}%
{2(1-p^{-1-s_{23}})}.
\end{align*}
Now, by using that $p\equiv3\operatorname{mod}4$, $\tau\neq\varepsilon$, and
Formula $2$, the contribution of all \ these integrals is
\begin{equation}
\frac{p^{-1-s_{23}}(1-p^{-1})}{2(1-p^{-1-s_{23}})}\frac{1}{p}\sum
\limits_{j=2}^{p-1}H_{\tau}(j)H_{\tau}(1-j)=\left(  \frac{p-3}{8p}\right)
\frac{p^{-1-s_{23}}(1-p^{-1})}{(1-p^{-1-s_{23}})}. \label{Formula_H_ij}%
\end{equation}

\textbf{Case} $i=1$ \textbf{and} $j=0$.

In this case by using Formula $1$,
\begin{align}
L_{10}(\boldsymbol{s},\tau)  &  =\int_{1+p\mathbb{Z}_{p}\times p\mathbb{Z}%
_{p}}|1-x_{2}|_{p}^{s_{42}}|x_{3}|_{p}^{s_{13}}H_{\tau}(1-x_{2})H_{\tau}%
(x_{3})dx_{2}dx_{3}\nonumber\\
&  =\left\{  \int_{1+p\mathbb{Z}_{p}}|1-x_{2}|_{p}^{s_{42}}H_{\tau}%
(1-x_{2})dx_{2}\right\}  \left\{  \int_{p\mathbb{Z}_{p}}|x_{3}|_{p}^{s_{13}%
}H_{\tau}(x_{3})dx_{3}\right\} \nonumber\\
&  =\left\{  \int_{p\mathbb{Z}_{p}}|x_{2}|_{p}^{s_{42}}H_{\tau}(-x_{2}%
)dx_{2}\right\}  \left\{  \int_{p\mathbb{Z}_{p}}|x_{3}|_{p}^{s_{13}}H_{\tau
}(x_{3})dx_{3}\right\} \nonumber\\
&  =\left\{  \frac{1}{2}\int_{p\mathbb{Z}_{p}}|x_{2}|_{p}^{s_{42}}%
dx_{2}\right\}  \left\{  \frac{1}{2}\int_{p\mathbb{Z}_{p}}|x_{3}|_{p}^{s_{13}%
}dx_{3}\right\}  =\frac{\left(  1-p^{-1}\right)  ^{2}}{4}\frac{p^{-2-s_{42}%
-s_{13}}}{\left(  1-p^{-1-s_{42}}\right)  \left(  1-p^{-1-s_{13}}\right)  }.
\label{Formula_L_01}%
\end{align}
\textbf{Case} $i=0$ \textbf{and} $j=1$.

Since
\[
H_{\tau}(x_{2}-x_{3})\Big\vert_{p\mathbb{Z}_{p}\times1+p\mathbb{Z}_{p}%
}=H_{\tau}(-1)=0,
\]
we have $\ L_{01}(\boldsymbol{s},\tau)=0$.

\textbf{Case }$i=j=0$.

In this case,%
\[
L_{00}(\boldsymbol{s},\tau)=\int_{\left(  p\mathbb{Z}_{p}\right)  ^{2}}%
|x_{2}|_{p}^{s_{12}}|x_{3}|_{p}^{s_{13}}|x_{2}-x_{3}|_{p}^{s_{23}}H_{\tau
}(x_{2})H_{\tau}(x_{3})H_{\tau}(x_{2}-x_{3})dx_{2}dx_{3}.
\]
By using that
\begin{gather*}
H_{\tau}(x_{2})H_{\tau}(x_{3})H_{\tau}(x_{2}-x_{3})=\frac{1}{8}\left\{
1+\mathrm{sgn}_{\tau}(x_{2})+\mathrm{sgn}_{\tau}(x_{3})+\mathrm{sgn}_{\tau
}(x_{2}-x_{3})\right. \\
+\mathrm{sgn}_{\tau}(x_{2})\mathrm{sgn}_{\tau}(x_{3})+\mathrm{sgn}_{\tau
}(x_{2})\mathrm{sgn}_{\tau}(x_{2}-x_{3})+\mathrm{sgn}_{\tau}(x_{3}%
)\mathrm{sgn}_{\tau}(x_{2}-x_{3})\\
\left.  +\mathrm{sgn}_{\tau}(x_{2})\mathrm{sgn}_{\tau}(x_{3})\mathrm{sgn}%
_{\tau}(x_{2}-x_{3})\right\}  ,
\end{gather*}
and the notation introduced in Formulae $6$ to $9$, we have%
\begin{align}
L_{00}(\boldsymbol{s},\tau)  &  =L_{00}(s_{12},s_{13},s_{23})+L_{00}^{\left(
1\right)  }(s_{12},s_{13},s_{23},\tau)+L_{00}^{\left(  2\right)  }%
(s_{12},s_{13},s_{23},\tau)+L_{00}^{\left(  3\right)  }(s_{12},s_{13}%
,s_{23},\tau)\nonumber\\
&  =L_{00}(s_{12},s_{13},s_{23})+L_{00}^{\left(  1\right)  }(s_{12}%
,s_{13},s_{23},\tau)+L_{00}^{\left(  1\right)  }(s_{12},s_{23},s_{13}%
,\tau)-L_{00}^{\left(  1\right)  }\left(  s_{13},s_{23},s_{12},\tau\right)  ,
\label{Formula_L_00_tau}%
\end{align}
the integrals \ involving an odd number of sign functions vanish. This fact
can be established by a suitable change of variables as in Formula $1$.

\textbf{Case }$i=j=1$.

In this case,%
\[
L_{11}(\boldsymbol{s},\tau)=\int_{\left(  1+p\mathbb{Z}_{p}\right)  ^{2}%
}|1-x_{2}|_{p}^{s_{42}}|1-x_{3}|_{p}^{s_{43}}|x_{2}-x_{3}|_{p}^{s_{23}}%
H_{\tau}(1-x_{2})H_{\tau}(1-x_{3})H_{\tau}(x_{2}-x_{3})dx_{2}dx_{3}.
\]
Now by changing variables as $u=1-x_{2}$, $v=1-x_{3}$, we get%
\begin{align}
L_{11}(\boldsymbol{s},\tau)  &  =\int_{\left(  p\mathbb{Z}_{p}\right)  ^{2}%
}|u|_{p}^{s_{42}}|v|_{p}^{s_{43}}|u-v|_{p}^{s_{23}}H_{\tau}(u)H_{\tau
}(v)H_{\tau}(v-u)dudv\nonumber\\
&  =L_{00}(s_{42},s_{43},s_{23})+L_{00}^{\left(  1\right)  }(s_{42}%
,s_{43},s_{23},\tau)-L_{00}^{\left(  2\right)  }(s_{42},s_{43},s_{23}%
,\tau)-L_{00}^{\left(  3\right)  }(s_{42},s_{43},s_{23},\tau)\nonumber\\
&  =L_{00}(s_{42},s_{43},s_{23})+L_{00}^{\left(  1\right)  }(s_{42}%
,s_{43},s_{23},\tau)-L_{00}^{\left(  1\right)  }\left(  s_{42},s_{23}%
,s_{43},\tau\right)  +L_{00}^{\left(  1\right)  }\left(  s_{43},s_{23}%
,s_{42},\tau\right)  \label{Formula_L_00_tau_1}%
\end{align}

\textbf{Cases }$i=0$\textbf{ and }$j\in\left\{  2,3,\ldots,p-1\right\}  $
\textbf{or} $i\in\left\{  2,3,\ldots,p-1\right\}  $ \textbf{and} $j=1$.

In these cases,
\begin{equation}
L_{0j}(\boldsymbol{s},\tau)=L_{i1}(\boldsymbol{s},\tau)=0.
\label{Formulas_L_0i_1i}%
\end{equation}

The vanishing of the integral $L_{0j}(\boldsymbol{s},\tau)$ follows from
\[
H_{\tau}(x_{3})H_{\tau}(x_{2}-x_{3})\Big\vert_{p\mathbb{Z}_{p}\times
j+p\mathbb{Z}_{p}}=H_{\tau}(j)H_{\tau}(-j)=0.
\]
The other case is treated in a similar way.

\textbf{Case }$i\in\left\{  2,3,\ldots,p-1\right\}  $\textbf{ and }$j=0$.

By using (\ref{Formula_I_0}),%
\begin{align}
L_{i0}(\boldsymbol{s},\tau)  &  =H_{\tau}^{2}(i)H_{\tau}(1-i)H_{\tau}%
(1)\int_{i+p\mathbb{Z}_{p}\times p\mathbb{Z}_{p}}|x_{3}|_{p}^{s_{13}}H_{\tau
}(x_{3})dx_{2}dx_{3}\nonumber\\
&  =p^{-1}H_{\tau}(i)H_{\tau}(1-i)\int_{p\mathbb{Z}_{p}}|x_{3}|_{p}^{s_{13}%
}H_{\tau}(x_{3})dx_{3}=p^{-1}H_{\tau}(i)H_{\tau}(1-i)I_{0}(s_{13}%
,\tau)\nonumber\\
&  =p^{-1}H_{\tau}(i)H_{\tau}(1-i)\frac{p^{-1-s_{13}}(1-p^{-1})}%
{2(1-p^{-1-s_{13}})}. \label{Formula_L_i0}%
\end{align}

Now, using Formula $2$, the contribution of all these integrals is
\begin{equation}
\sum\limits_{i=2}^{p-1}L_{i0}(\boldsymbol{s},\tau)=\frac{p^{-1-s_{13}%
}(1-p^{-1})}{2(1-p^{-1-s_{13}})}\frac{1}{p}\sum\limits_{i=2}^{p-1}H_{\tau
}(i)H_{\tau}(1-i)=\left(  \frac{p-3}{8p}\right)  \frac{p^{-1-s_{13}}%
(1-p^{-1})}{(1-p^{-1-s_{13}})}. \label{Formula_L_i0_A}%
\end{equation}

\textbf{Case }$i=1$ \textbf{and }$j\in\left\{  2,3,\ldots,p-1\right\}  $.

This case is similar to the previous one,
\begin{equation}
\sum\limits_{j=2}^{p-1}L_{1j}(\boldsymbol{s},\tau)=p^{-1}I_{0}(s_{42}%
,\tau)\sum\limits_{j=2}^{p-1}H_{\tau}(j)H_{\tau}(1-j)=\left(  \frac{p-3}%
{8p}\right)  \frac{p^{-1-s_{42}}(1-p^{-1})}{(1-p^{-1-s_{42}})}.
\label{Formula_L_1j}%
\end{equation}

In conclusion, from (\ref{Zeta_5}), (\ref{Zeta_5A}), and (\ref{Formula_L_ij})-
(\ref{Formula_L_1j}), we have%
\begin{gather}
Z^{(5)}(\boldsymbol{s},\widetilde{\boldsymbol{s}})=E^{(5)}(\widetilde
{\boldsymbol{s}})\left\{  \frac{(p-3)(p-7)}{32p^{2}}+\left(  \frac{p-3}%
{8p}\right)  \left[  \frac{p^{-1-s_{23}}(1-p^{-1})}{(1-p^{-1-s_{23}})}\right.
\right. \nonumber\\
\left.  +\frac{p^{-1-s_{13}}(1-p^{-1})}{(1-p^{-1-s_{13}})}+\frac{p^{-1-s_{42}%
}(1-p^{-1})}{(1-p^{-1-s_{42}})}\right]  +\frac{\left(  1-p^{-1}\right)  ^{2}%
}{4}\frac{p^{-2-s_{13}-s_{42}}}{\left(  1-p^{-1-s_{13}}\right)  \left(
1-p^{-1-s_{42}}\right)  }\nonumber\\
+\frac{1}{4}\frac{p^{-s_{12}-s_{13}-s_{23}-2}\left(  1-p^{-1}\right)
}{1-p^{-s_{12}-s_{13}-s_{23}-2}}\left[  \frac{1}{2}-\frac{3}{2p}%
+\frac{p^{-1-s_{23}}\left(  1-p^{-1}\right)  }{1-p^{-1-s_{23}}}+\frac
{p^{-1-s_{13}}\left(  1-p^{-1}\right)  }{1-p^{-1-s_{13}}}\right] \nonumber\\
\left.  +\frac{1}{4}\frac{p^{-s_{42}-s_{43}-s_{23}-2}\left(  1-p^{-1}\right)
}{1-p^{-s_{42}-s_{43}-s_{23}-2}}\left[  \frac{1}{2}-\frac{3}{2p}%
+\frac{p^{-1-s_{23}}\left(  1-p^{-1}\right)  }{1-p^{-1-s_{23}}}+\frac
{p^{-1-s_{42}}\left(  1-p^{-1}\right)  }{1-p^{-1-s_{42}}}\right]  \right\}  .
\label{cincopuntos}%
\end{gather}
$Z^{(5)}(\boldsymbol{s},\widetilde{\boldsymbol{s}})$ is a holomorphic function
in {\normalsize
\begin{align*}
&  \mathrm{Re}(s_{13})>-1;\quad\mathrm{Re}(s_{23})>-1;\quad\mathrm{Re}%
(s_{42})>-1;\\
&  \mathrm{Re}(s_{12}+s_{13}+s_{23})>-2;\quad\mathrm{Re}(s_{42}+s_{43}%
+s_{23})>-2.
\end{align*}
}

\section{\label{Section_VIII}The limit $p\rightarrow1$ of the Ghoshal-Kawano
amplitudes}

In \cite{Bocardo-Gaspar:2017atv} we established that the limit $p$ approaches
to one of $p$-adic open string amplitudes at the tree-level can be defined
rigorously by using the Denef and Loeser theory of topological zeta functions
\cite{denefandloeser}. Notice that the calculations involving the limit
$p\rightarrow1$ in the case of the effective action are performed in
$\mathbb{R}^{D}$, meanwhile the calculations involving the limit
$p\rightarrow1$ in the case of $p$-adic string amplitudes are performed in
$\mathbb{Q}_{p}^{D}$, and in the $p$-adic topology the limit $p\rightarrow1$
does not make sense. However,\ surprisingly, the computation of the limit
$p\rightarrow1$ (considering $p$ as a real parameter) of the $p$-adic open
string amplitudes gives the right answer! In this subsection we compute limit
$p\rightarrow1$ (considering $p$ as a real parameter) \ in the cases $N=4$,
$5$. The computation of the limit $p\rightarrow1$ in the general case require
the so called explicit formulas, see \cite{Bocardo-Gaspar:2017atv} for further details.

The limit $p\rightarrow1$ of $Z^{\left(  N\right)  }(\boldsymbol{s}%
,\widetilde{\boldsymbol{s}},\tau)$, $N=4$, $5$, with $p\equiv3$
$\operatorname{mod}4$, are given by
\begin{equation}
\lim_{p\rightarrow1}Z^{\left(  4\right)  }(\boldsymbol{s},\widetilde
{\boldsymbol{s}},\tau)=\exp\left\{  \frac{\sqrt{-1}}{2}\left(  \widetilde
{s}_{13}+\widetilde{s}_{12}+\widetilde{s}_{23}\right)  \right\}  \left\{
-\frac{1}{2}+\frac{1}{2(s_{12}+1)}+\frac{1}{2(s_{32}+1)}\right\}  ,
\label{cuatropuntosp1}%
\end{equation}
for $\tau\in\{p,\varepsilon p\}$, and
\begin{gather}
\lim_{p\rightarrow1}Z^{(5)}(\boldsymbol{s},\widetilde{\boldsymbol{s}}%
)=E^{(5)}(\widetilde{\boldsymbol{s}})\left\{  \frac{3}{16}-\frac{1}%
{4(s_{23}+1)}-\frac{1}{4(s_{13}+1)}-\frac{1}{4(s_{42}+1)}\right. \nonumber\\
+\frac{1}{4\left(  s_{42}+1\right)  \left(  s_{13}+1\right)  }+\frac
{1}{4\left(  s_{12}+s_{13}+s_{23}+2\right)  }\left[  -1+\frac{1}{(s_{23}%
+1)}+\frac{1}{(s_{13}+1)}\right] \nonumber\\
\left.  +\frac{1}{4\left(  s_{42}+s_{43}+s_{23}+2\right)  }\left[  -1+\frac
{1}{(s_{23}+1)}+\frac{1}{(s_{42}+1)}\right]  \right\}  . \label{cincopuntosp1}%
\end{gather}
In the case $N=4$, after the appropriate sum over the permutations of the
momenta $\boldsymbol{k}_{i}$, the amplitude agrees with the Feynman amplitude
obtained from the noncommutative version of the Gerasimov-Shatashvili action
with a logarithmic potential (\ref{linearaction}).

\section{\label{Section_IX}Final Remarks}

In the present article, we study the Ghoshal-Kawano amplitudes for $p$-adic
open strings at tree-level level \cite{Ghoshal:2004ay}. These amplitudes
include Chan-Paton factors and an external $B$-field.

In section \ref{Section_II} starting from the noncommutative effective action
(\ref{goodaction}) discussed in \cite{Ghoshal:2004dd,Ghoshal:2004ay}, in the
present article, we obtain the corresponding tree-level four-point amplitudes
(\ref{cuatropuntos}) in the limit $p\rightarrow1$. This result was achieved by
adapting the heuristic approach given in \cite{Gerasimov:2000zp} for the
noncommutative case. By an explicit computation using the noncommutative field
theory \cite{Minwalla:1999px,Micu:2000xj}, we determine the four-point
amplitude at the tree level coming from the noncommutative
Gerasimov-Shatashvili Lagrangian. This amplitude is the sum of the expressions
(\ref{vertex}) and (\ref{canalesstu}). The first one represents the
noncommutative vertex four-point function and the second one is the
superposition of the amplitudes corresponding to the noncommutative channels
$s$, $t$ and $u$. The calculated tree-level amplitude is completely described
by planar Feynman diagrams and consequently the noncommutativity effect arises
as a global phase factor in front of the amplitude. Five-point amplitudes (or
higher-order amplitudes) can be also computed in a straightforward way
following the same procedure.

The study of the $p$-adic Ghoshal-Kawano amplitudes requires the use of
multivariate local zeta functions involving multiplicative characters and a
phase factor including the noncommutative parameter $\theta$. These are new
mathematical objects. We call these objects Ghoshal-Kawano zeta functions. In
sections \ref{Section_IV} and \ref{Section_V}, by using Hironaka's resolution
of singularities theorem, we prove that these integrals admit meromorphic
continuation as complex functions in the external momenta of the $N$ external particles.

Four and five point amplitudes were computed explicitly in sections
\ref{Section_VI} and \ref{Section_VII}, see (\ref{cuatropuntos}) and
(\ref{cincopuntos}). The four-point amplitude (\ref{cuatropuntos}) coincides
with the one obtained in \cite{Ghoshal:2004ay}. The five-point amplitude was
not obtained previously.

In section \ref{Section_VIII} we study again the amplitudes from the
worldsheet view point. We compute the limit $p\rightarrow1$ for four and five
point amplitudes resulting in the formulae (\ref{cuatropuntosp1}) and
(\ref{cincopuntosp1}), respectively. The four-point amplitude
(\ref{cuatropuntosp1}) agrees with the heuristic computation given by the
superposition of formulae (\ref{vertex}) and (\ref{canalesstu}).

As we mentioned before, in the computation of Ghoshal-Kawano amplitudes at the
tree-level, the noncommutative effect coming from the constant $B$-field
arises only as a global phase factor because only planar diagrams are
involved. In the computation of amplitudes at one-loop or multi-loops,
non-planar diagrams systematically arise. It would be very interesting to
study the possibility of finding a non-trivial noncommutative effect, as the
IR/UV mixing, as a result of the contribution of one-loop non-planar diagrams.
Probably the multi-loop analysis of the $p$-adic string theory studied in
\cite{Chekhov:1989bg}, will play an important role for the analysis of the
IR/UV mixing and other interesting effects of the $B$-field in $p$-adic string
theory amplitudes.

On the other hand, we think that the study of the amplitudes (\ref{Amplitude})
\ without the ad hoc normalization $x_{1}=0$, $x_{N-1}=1$, $x_{N}=\infty$ may
provide new insights on the effects of the $B$-field in $p$-adic string theory
amplitudes. However, the study of these amplitudes is more involved than the
one done here. Some of this work is in progress and will be reported elsewhere.

\section{\label{Section_Appendix}Appendix: basic aspects of the $p$-adic
analysis}

In this appendix we collect some basic results about $p$-adic analysis that
will be used in the article. For an in-depth review of the $p$-adic analysis
the reader may consult \cite{Alberio et al,Taibleson,V-V-Z}.

\subsection{The field of $p$-adic numbers}

Along this article $p$ will denote a prime number different from $2$. The
field of $p-$adic numbers $%
\mathbb{Q}
_{p}$ is defined as the completion of the field of rational numbers
$\mathbb{Q}$ with respect to the $p-$adic norm $|\cdot|_{p}$, which is defined
as
\[
\left\vert x\right\vert _{p}=\left\{
\begin{array}
[c]{lll}%
0 & \text{if} & x=0\\
&  & \\
p^{-\gamma} & \text{if} & x=p^{\gamma}\frac{a}{b}\text{,}%
\end{array}
\right.
\]
where $a$ and $b$ are integers coprime with $p$. The integer $\gamma:=ord(x)$,
with $ord(0):=+\infty$, is called the\ $p-$adic order of $x$.

Any $p-$adic number $x\neq0$ has a unique expansion of the form
\[
x=p^{ord(x)}\sum_{j=0}^{\infty}x_{j}p^{j},
\]
where $x_{j}\in\{0,\dots,p-1\}$ and $x_{0}\neq0$. In addition, any non-zero
$p-$adic number can be represented uniquely as $x=p^{ord(x)}ac\left(
x\right)  $ where $ac\left(  x\right)  =\sum_{j=0}^{\infty}x_{j}p^{j}$,
$x_{0}\neq0$, is called the angular component of $x$. Notice that $\left\vert
ac\left(  x\right)  \right\vert _{p}=1$.

We extend the $p-$adic norm to $%
\mathbb{Q}
_{p}^{n}$ by taking
\[
||x||_{p}:=\max_{1\leq i\leq n}|x_{i}|_{p},\text{ for }x=(x_{1},\dots
,x_{n})\in%
\mathbb{Q}
_{p}^{n}.
\]
We define $ord(x)=\min_{1\leq i\leq n}\{ord(x_{i})\}$, \ then \ $||x||_{p}%
=p^{-ord(x)}$.\ The metric space $\left(
\mathbb{Q}
_{p}^{n},||\cdot||_{p}\right)  $ is a separable complete ultrametric space.
For $r\in\mathbb{Z}$, denote by $B_{r}^{n}(a)=\{x\in%
\mathbb{Q}
_{p}^{n};||x-a||_{p}\leq p^{r}\}$ the ball of radius $p^{r}$ with center at
$a=(a_{1},\dots,a_{n})\in%
\mathbb{Q}
_{p}^{n}$, and take $B_{r}^{n}(0):=B_{r}^{n}$. Note that $B_{r}^{n}%
(a)=B_{r}(a_{1})\times\cdots\times B_{r}(a_{n})$, where $B_{r}(a_{i}):=\{x\in%
\mathbb{Q}
_{p};|x_{i}-a_{i}|_{p}\leq p^{r}\}$ is the one-dimensional ball of radius
$p^{r}$ with center at $a_{i}\in%
\mathbb{Q}
_{p}$. The ball $B_{0}^{n}$ equals the product of $n$ copies of $B_{0}%
=\mathbb{Z}_{p}$, the ring of $p-$adic integers of $%
\mathbb{Q}
_{p}$. We also denote by $S_{r}^{n}(a)=\{x\in Q_{p}^{n};||x-a||_{p}=p^{r}\}$
the sphere of radius $p^{r}$ with center at $a=(a_{1},\dots,a_{n})\in%
\mathbb{Q}
_{p}^{n}$, and take $S_{r}^{n}(0):=S_{r}^{n}$. We notice that $S_{0}%
^{1}=\mathbb{Z}_{p}^{\times}$ (the group of units of $\mathbb{Z}_{p}$), but
$\left(  \mathbb{Z}_{p}^{\times}\right)  ^{n}\subsetneq S_{0}^{n}$. The balls
and spheres are both open and closed subsets in $%
\mathbb{Q}
_{p}^{n}$. In addition, two balls in $%
\mathbb{Q}
_{p}^{n}$ are either disjoint or one is contained in the other.

As a topological space $\left(
\mathbb{Q}
_{p}^{n},||\cdot||_{p}\right)  $ is totally disconnected, i.e. the only
connected \ subsets of $%
\mathbb{Q}
_{p}^{n}$ are the empty set and the points. A subset of $%
\mathbb{Q}
_{p}^{n}$ is compact if and only if it is closed and bounded in $%
\mathbb{Q}
_{p}^{n}$, see e.g. \cite[Section 1.3]{V-V-Z}, or \cite[Section 1.8]{Alberio
et al}. The balls and spheres are compact subsets. Thus $\left(
\mathbb{Q}
_{p}^{n},||\cdot||_{p}\right)  $ is a locally compact topological space.

\subsection{Integration}

Since $(\mathbb{Q}_{p},+)$ is a locally compact topological group, there
exists a measure $dx$, which is invariant under translations, i.e.
$d(x+a)=dx$. If we normalize this measure by the condition $\int
_{\mathbb{Z}_{p}}dx=1$, then $dx$ is unique. A such measure is called the Haar
measure of $(\mathbb{Q}_{p},+)$. In the $n$-dimensional case, $(\mathbb{Q}%
_{p}^{n},+)$ is also a locally compact topological group. We denote by
$d^{n}x$ the Haar measure normalized by the condition $\int_{\mathbb{Z}%
_{p}^{n}}d^{n}x=1$. This measure agrees with the product measure $dx_{1}\cdots
dx_{n}$, and it also satisfies that $d^{n}(x+a)=d^{n}x$, for $a\in
\mathbb{Q}_{p}^{n}$

A function $h:U\rightarrow\mathbb{Q}_{p}$ is said to be analytic on an open
subset $U\subseteq\mathbb{Q}_{p}^{n}$, if for every $b=(b_{1},\ldots,b_{n})\in
U$ there exists an open subset $\widetilde{U}\subset U$, with $b\in
\widetilde{U}$, and a convergent power series $\sum_{i\in\mathbb{N}^{n}}%
a_{i}\left(  x-b\right)  ^{i}$ for $x=(x_{1},\ldots,x_{n})\in$ $\widetilde{U}%
$, such that $h\left(  x\right)  =\sum_{i\in\mathbb{N}^{n}}a_{i}\left(
x-b\right)  ^{i}$ for $x\in\widetilde{U}$, with \ $i=(i_{1},\ldots,i_{n})$ and
$\left(  x-b\right)  ^{i}=\prod\nolimits_{j=1}^{n}\left(  x_{j}-b_{j}\right)
^{i_{j}}$. In this case, $\frac{\partial}{\partial x_{l}}h\left(  x\right)
=\sum_{i\in\mathbb{N}^{n}}a_{i}\frac{\partial}{\partial x_{l}}\left(
x-b\right)  ^{i}$ is a convergent power series.

Let $U$, $V$ be open subsets in $\mathbb{Q}_{p}^{n}$. A mapping
$H:U\rightarrow V$, $H=(H_{1},\ldots,H_{n})$ is called analytic if each
$H_{i}$\ is analytic. The mapping $H$ is said to be bi-analytic if $H$ and
$H^{-1}$ are analytic.

\subsubsection{Change of variables formula}

Let $K_{0},K_{1}\subset\mathbb{Q}_{p}^{n}$ be open compact subsets, and let
$H=\left(  H_{1},\ldots,H_{n}\right)  :K_{1}\rightarrow K_{0}$ be a
bi-analytic map such that
\[
\det\left[  \frac{\partial H_{i}}{\partial y_{j}}(y)\right]  \neq0\text{, for
\ }y\in K_{1}.
\]
If $f$ is a continuous function on $K_{0}$, then
\[%
{\textstyle\int\limits_{K_{0}}}
f\left(  x\right)  \ d^{n}x=%
{\textstyle\int\limits_{K_{1}}}
f\left(  H(y)\right)  \left\vert \det\left[  \frac{\partial H_{i}}{\partial
y_{j}}(y)\right]  \right\vert _{p}\ d^{n}y,\quad\text{ \ (}x=H(y)\text{).}%
\]
For the proof of this theorem the reader may consult \cite[Prop. 7.4.1]{Igusa}
or \cite[Section 10.1.2]{Bourbaki}.

\subsection{Some arithmetic functions}

In this section we review some arithmetic functions that we shall use
throughout this article.

\subsubsection{Multiplicative characters}

A multiplicative character (or quasi-character) of the group $\left(
\mathbb{Q}_{p}^{\times},\cdot\right)  $ is a continuous homomorphism
$\chi:\mathbb{Q}_{p}^{\times}\rightarrow\mathbb{C}^{\times}$ satisfying
$\chi\left(  xy\right)  =\chi\left(  x\right)  \chi\left(  y\right)  $. Every
multiplicative character has the form%
\[
\chi\left(  x\right)  =\left\vert x\right\vert _{p}^{s}\chi_{0}\left(
ac(x)\right)  \text{, for some }s\in\mathbb{C}\text{,}%
\]
where $\chi_{0}$ is the restriction of $\chi$ to $\mathbb{Z}_{p}^{\times}$,
which is a continuous multiplicative character of $\left(  \mathbb{Z}%
_{p}^{\times},\cdot\right)  $ into the complex unit circle.

\subsubsection{The Legendre symbol}

For $a$ an integer number and $p$ a prime number, the Legendre symbol is
defined as%
\[
\bigg(\frac{a}{p}\bigg)=\left\{
\begin{array}
[c]{lll}%
1 & \text{if} & x^{2}\equiv a\ mod\ p\ \text{has\ a\ solution}\\
&  & \\
-1 & \text{if} & \text{otherwise.}%
\end{array}
\right.
\]
The following formulas are used in several calculations in this article:%
\[
\bigg(\frac{1}{p}\bigg)=1;\hspace{0.5cm}\text{ }\bigg(\frac{-1}{p}%
\bigg)=(-1)^{\frac{p-1}{2}}=\left\{
\begin{array}
[c]{lll}%
1 & \text{if} & p\equiv1\ mod\ 4\\
&  & \\
-1 & \text{if} & p\equiv3\ mod\ 4.
\end{array}
\right.
\]
Take for $x\in\mathbb{Q}_{p}^{\times}$, $ac(x)=x_{0}+x_{1}p+\ldots
\in\mathbb{Z}_{p}^{\times}$, then
\[%
\begin{array}
[c]{lll}%
\mathbb{Q}_{p}^{\times} & \rightarrow & \left\{  \pm1\right\} \\
&  & \\
x & \rightarrow & (\frac{x_{0}}{p})
\end{array}
\]
is a unitary multiplicative character.

\subsubsection{The sign function}

We review the $p$-adic sign function, which is a multiplicative character with
values in $\left\{  \pm1\right\}  $. We denote $\left[  \mathbb{Q}_{p}%
^{\times}\right]  ^{2}$ the multiplicative subgroup of squares in
$\mathbb{Q}_{p}^{\times}$, i.e.
\[
\left[  \mathbb{Q}_{p}^{\times}\right]  ^{2}=\{a\in\mathbb{Q}_{p}%
;a=b^{2}\text{ for some }b\in\mathbb{Q}_{p}^{\times}\}.
\]
For $p\neq2$, and $\varepsilon\in\{1,\ldots,p-1\}$ satisfying $(\frac
{\varepsilon}{p})=-1$, we have
\[
\mathbb{Q}_{p}^{\times}/\left[  \mathbb{Q}_{p}^{\times}\right]  ^{2}%
=\{1,\varepsilon,p,\varepsilon p\},
\]
This means that any nonzero $p$-adic number can be written uniquely as
\[
x=\tau a^{2},\text{ with }a\in\mathbb{Q}_{p}^{\times}\text{ and }\tau
\in\mathbb{Q}_{p}^{\times}/\left[  \mathbb{Q}_{p}^{\times}\right]  ^{2}.
\]
For a fixed $\tau\in\{\varepsilon,p,\varepsilon p\}$, and $x\in\mathbb{Q}%
_{p}^{\times}$, we set
\begin{equation}
\mathrm{sgn}_{\tau}(x):=%
\begin{cases}
1 & \text{if}\ x=a^{2}-\tau b^{2}\ \text{for}\ a,b\in\mathbb{Q}_{p}\\
-1 & \text{otherwise.}%
\end{cases}
\label{signdef}%
\end{equation}
The following is the list of all the possible $p$-adic sign functions:%
\begin{equation}%
\begin{tabular}
[c]{|l|l|}\hline
$p\equiv1$ $\operatorname{mod}$ $4$ & $p\equiv3$ $\operatorname{mod}$
$4$\\\hline
$\mathrm{sgn}_{\varepsilon}(x)=\left(  -1\right)  ^{ord(x)}$ & $\mathrm{sgn}%
_{\varepsilon}(x)=\left(  -1\right)  ^{ord(x)}$\\\hline
$\mathrm{sgn}_{p}(x)=\left(  \frac{x_{0}}{p}\right)  $ & $\mathrm{sgn}%
_{p}(x)=\left(  -1\right)  ^{ord(x)}\left(  \frac{x_{0}}{p}\right)  $\\\hline
$\mathrm{sgn}_{\varepsilon p}(x)=\left(  -1\right)  ^{ord(x)}\left(
\frac{x_{0}}{p}\right)  $ & $\mathrm{sgn}_{\varepsilon p}(x)=\left(
\frac{x_{0}}{p}\right)  ,$\\\hline
\end{tabular}
\ \label{Table}%
\end{equation}
see \cite{Gubser:2018cha}. \ Then $\mathrm{sgn}_{\tau}$ is a multiplicative
character, and a locally constant function in $\mathbb{Q}_{p}^{\times}$, more
precisely, $\mathrm{sgn}_{\tau}(x-y)=\mathrm{sgn}_{\tau}(x)$ if
$ord(y)>ord(x)$.

We take $p\equiv3$ $\operatorname{mod}$ $4$ and $\tau\in\left\{  p,\varepsilon
p\right\}  $, to have $\mathrm{sgn}_{\tau}(-y)=-\mathrm{sgn}_{\tau}(y)$, for
any $y\in\mathbb{Q}_{p}^{\times}$. In all the calculations involving
$\mathrm{sgn}_{\tau}$ we assume that $p\equiv3$ $\operatorname{mod}$ $4$ and
$\tau\in\left\{  p,\varepsilon p\right\}  $. We define the Heaviside step
function as
\begin{equation}
H_{\tau}(x):=H_{\tau}^{+}(x)=\frac{1}{2}(1+\mathrm{sgn}_{\tau}(x))=\left\{
\begin{array}
[c]{ll}%
1 & \text{if \textrm{sgn}$_{\tau}(x)=1$}\\
& \\
0 & \text{if \textrm{sgn}$_{\tau}(x)=-1,$}%
\end{array}
\right.  \label{Haviside}%
\end{equation}
for any $x\in\mathbb{Q}_{p}^{\times}$. It is convenient to set
\[
H_{\tau}^{\pm}(x):=\frac{1}{2}(1\pm\mathrm{sgn}_{\tau}(x)).
\]
The following properties are useful:%
\begin{align*}
H_{\tau}(x)H_{\tau}(x) &  =H_{\tau}(x);\text{ \ }H_{\tau}(x)+H_{\tau}%
^{-}(x)=1;\text{ \ }H_{\tau}(x)H_{\tau}^{-}(x)=0;\text{ \ }\\
H_{\tau}(xy) &  =H_{\tau}(x)H_{\tau}(y)+\ H_{\tau}^{-}(x)H_{\tau}^{-}(y).
\end{align*}

\begin{acknowledgement}
The authors wish to thank the referees for their careful reading of the
original manuscript and for all the suggestions provided, which  helped us to
improve our paper.
\end{acknowledgement}

\end{document}